\documentclass[a4paper,fleqn]{cas-dc}

\usepackage[numbers]{natbib}

\usepackage[linesnumbered, vlined, ruled, commentsnumbered]{algorithm2e}
\usepackage{bm}
\usepackage[dvipsnames]{xcolor}

\usepackage{amsmath}
\usepackage{amsfonts,amssymb}
\usepackage{amsthm}
\usepackage[noend]{algpseudocode}
\usepackage{booktabs}
\usepackage{footnote}
\usepackage{colortbl}
\usepackage{tabularx}
\usepackage{multirow}
\usepackage{makecell}
\usepackage{threeparttable}
\usepackage{graphicx}
\graphicspath{{figs/}}
\usepackage{epstopdf}
\usepackage{subfigure}
\usepackage{hhline}
\usepackage{hyperref} % 这会让参考文献加下划线
\usepackage{longtable}
\usepackage{pifont}% http://ctan.org/pkg/pifont
\usepackage[normalem]{ulem} % 用于 \ulem，normalem 用于不给参考文献加下划线
\usepackage{graphbox}
\usepackage{cancel}
\usepackage{mathtools}
\usepackage{thmtools, thm-restate}
\usepackage{tikz}
\usepackage{paralist}
\usepackage{enumitem}
\usepackage{url}
\usepackage{xspace}
\usepackage{listings}
\usepackage{setspace}
\usepackage[skins,breakable,hooks]{tcolorbox}
\usepackage{float}
\usepackage{placeins}
\usepackage{array}
\usepackage{blindtext}
\usepackage{mdwlist}
\usepackage{color}
\usepackage{fancyhdr}


\usepackage{cryptocode}
\createpseudocodeblock{pcb}{center, boxed}{}{}{}

\newtheorem{theorem}{Theorem}

\newtheorem{definition}{Definition}

\definecolor{BrickRed}{RGB}{178,34,34}

\newcommand{\no}{\textcolor{Maroon}{\faTimes}}
\newcommand{\LEFTcircle}{%
  \tikz[baseline=-0.55ex,scale=0.12]{%
    \fill (0,1) arc[start angle=90,end angle=270,radius=1] -- cycle;
    \draw (0,0) circle[radius=1];}}
\newcommand{\CIRCLE}{%
  \tikz[baseline=-0.55ex,scale=0.12]{\fill (0,0) circle[radius=1];}}

\makeatletter  
\newif\if@restonecol  
\makeatother

\tcbset{%
  fancytitle/.code={%
    \ifstrempty{#1}{}{\pgfkeysalso{%
        top=\baselineskip,
        overlay unbroken and first app = {%
        \node[draw, rounded corners, line width=1pt,fill=white, anchor=west, xshift=5mm,inner xsep=2pt,inner ysep=2pt] at (frame.north west)%
        {\strut#1};},
       enlarge top by = \baselineskip,
       enlarge top at break by= 0mm,
    }}},
  fancybox/.style 2 args = {%
    breakable, rounded corners, enhanced, colback=white, fontupper=\footnotesize, % Change \small to \footnotesize here
    colframe=black, width=#1, toprule=1pt, rightrule=1pt,leftrule=1pt,bottomrule=1pt, left = 1.5pt,right=1.5pt, bottom = 1.5pt, fancytitle={#2},pad at break=2mm},
  fancybox/.default = {.5\textwidth}{}%

    fancy2title/.code={%
    \ifstrempty{#1}{}{\pgfkeysalso{%
        top=\baselineskip,
        overlay unbroken and first app = {%
        \node[draw, rounded corners, line width=1pt,fill=white, anchor=west, xshift=5mm,inner xsep=2pt,inner ysep=2pt] at (frame.north west)%
        {\strut#1};},
       enlarge top by = \baselineskip,
       enlarge top at break by= 0mm,
    }}},
  fancy2box/.style 2 args = {%
    rounded corners, enhanced, colback=white, fontupper=\footnotesize,
    colframe=black, width=#1, toprule=1pt, rightrule=1pt,leftrule=1pt,bottomrule=1pt, left = 1.5pt,right=1.5pt, bottom = 1.5pt, fancytitle={#2},pad at break=2mm},
  fancy2box/.default = {1\textwidth}{}%
}

\newtcolorbox{FancyBox}[2][.5\textwidth]{%
    fancybox={#1}{#2}}

\newcommand{\mylongbox}[4]{
\begin{FancyBox}[\linewidth]{#1}
\begin{flushleft}
        {\small #2} % Change \small to \footnotesize here
\end{flushleft}
\end{FancyBox}
\vspace{-15pt}
\begin{figure}[!h]
    \vspace{0pt}
    \caption{#3}
    \label{#4}
\end{figure}
}

\newcommand{\myonecolumnbox}[3]{
\begin{figure}[!ht]
\begin{FancyBox}[\linewidth]{#1}
\begin{flushleft}
        {\small #2}
\end{flushleft}
\end{FancyBox}
\vspace{-6pt}
    \caption{#3}
\end{figure}
\FloatBarrier
}

\newtcolorbox{Fancy2Box}[2][1\textwidth]{%
    fancy2box={#1}{#2}}
    
\newcommand{\mylongboxTwoColumn}[3]{
\begin{figure*}[!t]
\begin{Fancy2Box}[\textwidth]{#1}
\begin{flushleft}
        {\footnotesize #2}
\end{flushleft}
\end{Fancy2Box}
\vspace{-6pt}
    \caption{#3}
\end{figure*}
}

\def\BibTeX{{\rm B\kern-.05em{\sc i\kern-.025em b}\kern-.08em
    T\kern-.1667em\lower.7ex\hbox{E}\kern-.125emX}}

\newcommand{\heading}[1]{{\vspace{3pt}\noindent{\textbf{#1}}}}

\newenvironment{packeditemize}{
	\begin{list}{$\bullet$}{
			\setlength{\labelwidth}{4pt}
			\setlength{\itemsep}{0pt}
			\setlength{\leftmargin}{\labelwidth}
			\addtolength{\leftmargin}{\labelsep}
			\setlength{\parindent}{0pt}
			\setlength{\listparindent}{\parindent}
			\setlength{\parsep}{0pt}
			\setlength{\topsep}{1pt}}}{\end{list}}

\begin{document}
\let\WriteBookmarks\relax
\def\floatpagepagefraction{1}
\def\textpagefraction{.001}

\shorttitle{Starfish: Rebalancing Multi-Party Off-Chain Payment Channels}
\shortauthors{Yu et~al.}

\title[mode=title]{Starfish: Rebalancing Multi-Party Off-Chain Payment Channels}

% Provisional CRediT roles modeled on the published xRWA article.
% All authors should confirm these roles before submission.
\author[1]{Wenxuan Yu}
\credit{Writing -- original draft, Methodology, Software, Investigation}

\author[1]{Minghui Xu}
\ead{mhxu@sdu.edu.cn}
\cormark[1]
\credit{Supervision, Funding acquisition, Conceptualization, Writing -- review \& editing}

\author[2]{Yuanzhe Zhang}
% \ead{yuanzhe.zhang@ntu.edu.sg}
\credit{Formal analysis, Writing -- review \& editing}

\author[3]{Guangyong Shang}
\credit{Software, Validation}

\author[3]{Guangpeng Qi}
\credit{Software, Validation}

\author[1]{Dongliang Duan}
\credit{Investigation, Data curation}

\author[4]{Shan Wang}
\credit{Formal analysis, Writing -- review \& editing}

\author[1]{Kun Li}
\credit{Visualization, Validation}

\author[1]{Yue Zhang}
\credit{Supervision, Project administration, Writing -- review \& editing}

\cortext[1]{\hspace{0.25em}Corresponding author.}

\address[1]{School of Computer Science and Technology, Shandong University, Qingdao 266237, China}
\address[2]{Digital Trust Centre, Nanyang Technological University}
\address[3]{Inspur Yunzhou Industrial Internet Co., Ltd, Jinan 250101, China}
\address[4]{Department of Computing, The Hong Kong Polytechnic University, Hong Kong, China}

\begin{abstract}
Blockchain technology has revolutionized the way transactions are executed, but scalability remains a major challenge. Payment Channel Network (PCN), as a Layer-2 scaling solution, has been proposed to address this issue. However, skewed payments can deplete the balance of one party within a channel, restricting the ability of PCNs to transact through a path and subsequently reducing the transaction success rate.
To address this issue, the technology of rebalancing has been proposed.
However, existing rebalancing strategies in PCNs are limited in their efficiency and applicability under realistic network topologies. Cycle-based approaches only enable rebalancing among nodes that form cycles, while non-cycle-based approaches incur high on-chain operation costs.
In this study, we propose Starfish, a rebalancing approach designed for the locally star-like structures formed by high-degree routing hubs and their incident channels in real-world PCNs. 
By leveraging a hub-centric structure, Starfish reduces the on-chain overhead of rebalancing from quadratic to linear in the number of channels, achieving high rebalancing efficiency. 
We formally prove the security of Starfish and conduct comparative experiments against existing rebalancing techniques to demonstrate its effectiveness.
\end{abstract}


\maketitle

\section{Introduction}
Over the past decade, blockchain technology~\cite{chen2022survey} has experienced a rapid development as a trusted and decentralized means of executing transactions. However, its scalability remains a significant challenge. As a response, the research community and industry have proposed a number of Layer-2 scaling solutions to handle transactions off the Mainnet (Layer 1)~\cite{hearn2013micro, dziembowski2019perun, network2018cheap,xu2023trustless}. Payment channel is an innovative layer-2 scaling technique that enables transaction processing off-chain and final settlement on-chain. The collaboration of interconnected payment channels forms a payment channel network (PCN)~\cite{poon2016bitcoin,aumayr2021bitcoin, miller2019sprites, egger2019atomic, mccorry2019pisa, yu2022zk, guo2023cross}. 
% Within such a network, participants who do not own directly-connected payment channels can utilize paths formed by multiple channels to facilitate payments. There have been a few solution approaches~\cite{miller2019sprites, prihodko2016flare} to enhance the efficiency of a PCN, but the scalability issue still exists. 
However, the channel depletion problem, which refers to the scenario where a payment channel within the network runs out of funds, has a strong impact on the scalability of a PCN~\cite{li2020secure, luo2022learning, avarikioti2018payment}.
Such a phenomenon occurs when a significant number of transactions flow through a particular channel, exhausting its available balance. These transactions, known as skewed ones, render the channel temporarily unusable for further transactions until it is replenished with funds.
% Particularly, funds within channels in a PCN are limited, and payments are typically skewed, leading to a depletion of channel balances, consequently restricting additional payments within the channel.

Researchers have proposed two types of approaches to address the issue of channel depletion: routing-based and rebalancing-based.
Routing-based solutions~\cite{sivaraman2020high, yang2023optimal, wang2023fence, liu2023balanced, jiang2023balance, lin2020fstr} aim to balance and facilitate transactions within a payment channel network by selecting appropriate routing strategies. Rebalancing-based solutions~\cite{khalil2017revive,hong2022cycle,ni2023utility,ge2022shaduf}, on the other hand, realize the goal of rebalancing channels with depleted funds by transferring balances between channels. These two techniques complement each other, with the former helping to choose appropriate payment paths, while the latter reviving payment routes that were previously unusable. 
We focus on rebalancing-based methods in this paper and motivate their limitations from the perspective of real-world PCN topologies.

\begin{figure}[!htb]
\centering
\subfigure[]{
% \begin{minipage}[t]{0.25\linewidth}
\includegraphics[width=0.4\linewidth]{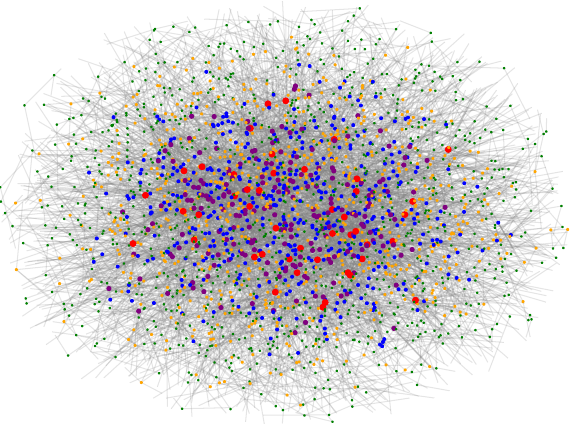}
\label{fig:lightning:topology}
% \end{minipage}%
}
\subfigure[]{
% \begin{minipage}[t]{0.25\linewidth}
\includegraphics[width=0.45\linewidth]{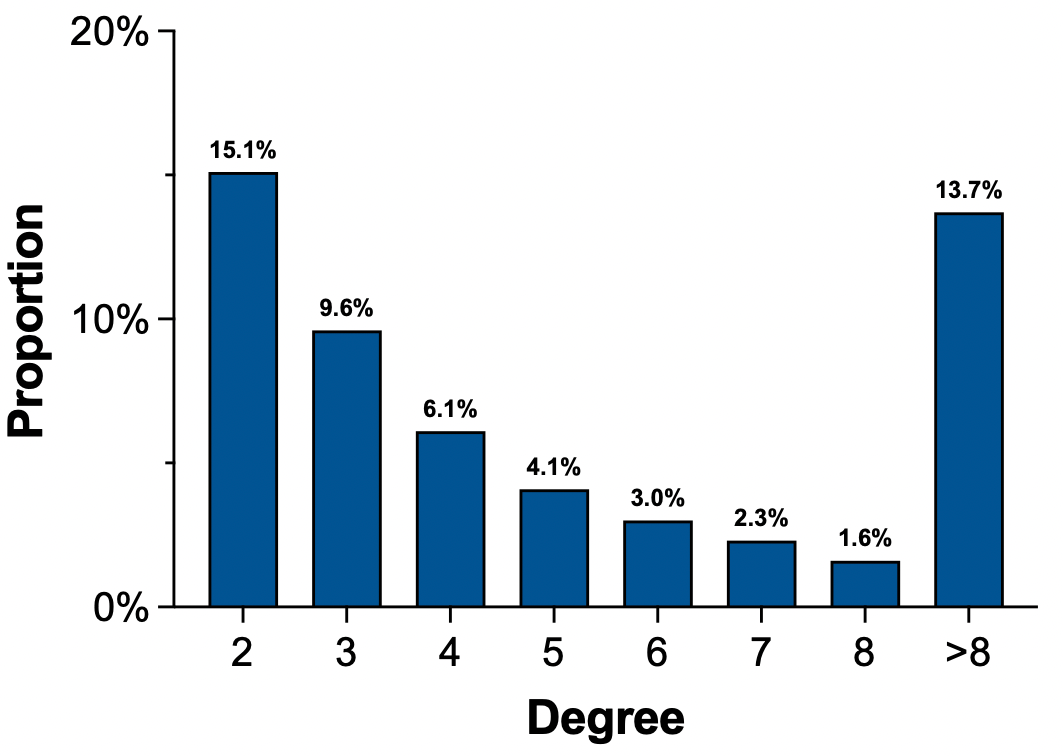}
\label{fig:dist}
% \end{minipage}%
}%
\centering
\caption{(a) Lightning Network visualization. Node colors indicate degree: white, 1; green, 2; orange, 3--4; blue, 5--8; purple, 9--16; and red, 17 or greater. Gray edges represent payment channels. (b) Node degree distribution, with degrees greater than 8 grouped as $>8$.}
\label{fig:lightning:topology:dist}
\end{figure}

% \begin{figure}[!t]
% \centering
% \subfigure[Storage utilization]{
% \label{fig:storage:usage}
% \includegraphics[width=0.48\linewidth]{./figures/geth_usage.pdf}}
% \subfigure[Storage utilization ratio]{
% \label{fig:storage:ratio}
% \includegraphics[width=0.48\linewidth]{./figures/geth_ratio.pdf}}
% \caption{Storage metrics of Ethereum geth node}
% \label{fig:storage}
% \end{figure}

%\textbf{Limitations of rebalancing-based approaches. }

A recent study~\cite{khamis2024topologies} shows that real-world PCNs are well captured by multi-star models, in which high-degree hubs connect many neighboring nodes. Accordingly, each high-degree hub and its incident channels form a locally star-like structure. In particular, both the Ethereum-based Raiden PCN~\cite{raiden2017} and the Bitcoin-based Lightning Network~\cite{poon2016bitcoin} exhibit power-law degree distributions consistent with the presence of high-degree hubs. 
Fig.~\ref{fig:lightning:topology:dist}(a) visualizes a snapshot of the Lightning Network topology\footnote{As of 15 October 2025, the Lightning Network comprised approximately 12,632 active nodes and 43,758 payment channels.}. 
Each gray edge represents a payment channel. 
In this topology analysis, intermediary nodes refer to nodes with degree at least two that can act as routing nodes. As shown in Fig.~\ref{fig:lightning:topology:dist}(b), approximately 70\% of intermediary nodes have degree greater than two. This indicates that many routing nodes have multiple incident channels, making the local structures around high-degree routing hubs suitable for Starfish.
High-degree nodes tend to carry more transaction flow, which increases their exposure to channel depletion under skewed payments. 
However, current rebalancing solutions are not well-suited for rebalancing among the incident channels of such hubs, limiting their ability to achieve high rebalancing efficiency and support large aggregate channel capacity.

% Fig.~\ref{fig:lightning:topology:dist}(a) presents a visual representation of the real-world Bitcoin lightning network\footnote{We get the snapshot of the lightning network topology on 2021-03-31, which contains 10,529 nodes and 38,910 channels.}, exhibiting its mesh topology. Each gray edge represents a payment channel. %Rebalancing operations aim to achieve equilibrium among two or more contiguous payment channels. 
% Fig.~\ref{fig:lightning:topology:dist}(b) illustrates the dense nature of the Bitcoin lightning network, wherein approximately 70\% of the intermediary nodes possess a degree exceeding two. Nodes characterized by high degrees may experience heightened transactional throughput, thereby elevating the risk of channel depletion in the presence of skewed transactions. However, such nodes also benefit from enhanced opportunities to effectively rebalance channels, consequently mitigating depletion risks. Current solutions are inadequately tailored for rebalancing within such a dense network topology, impeding their ability to achieve optimal rebalancing efficiency and accommodate large channel capacities. 

Two categories of rebalancing approaches exist: cycle-based \cite{khalil2017revive, hong2022cycle, ni2023utility, avarikioti2022hide} and non-cycle-based \cite{poon2016bitcoin, loop, ge2022shaduf}. The cycle-based approach, exemplified by the Revive protocol \cite{khalil2017revive}, leverages nodes organized in a cyclic network topology for rebalancing. Cycle \cite{hong2022cycle} further augments the efficacy of Revive. 
However, identifying and maintaining cycles incurs non-negligible time and bandwidth overhead. 
Moreover, when the channels incident to a routing hub do not form a cycle, cycle-based methods become inefficient or inapplicable.
Close-Open~\cite{poon2016bitcoin} and LOOP~\cite{loop} are two non-cycle-based solutions that achieve channel rebalancing by reopening a new channel. 
These methods are simple but require frequent on-chain interactions for every rebalancing transaction, leading to substantial latency and financial expenses.
In contrast, Shaduf~\cite{ge2022shaduf} introduces a different design: it performs a one-time on-chain binding between channels to establish a transferable relation, after which rebalancing can be executed repeatedly off-chain. 
However, in real-world PCNs where a hub connects to $N$ channels, enabling flexible rebalancing across all its incident channels requires pairwise binding among them, which entails $N(N-1)$ on-chain binding operations, making the approach poorly scalable for large hubs.

% Shaduf~\cite{ge2022shaduf}, as a non-cycle-based rebalancing approach, rebalances channels by binding (each costs 2$\times$ on-chain operations) adjacent channels and enabling balance transferring between them. When using the common ``All Bind'' or ``All to One'' binding strategies, the required number of on-chain operations to bind $N$ channels are respectively $N(N-1)$ or $2(N-1)$. Additionally, the channel capacity in Shaduf is limited by the highest capacity among the channels bound together.

In this paper, we present Starfish, a rebalancing protocol tailored to the locally star-like structure formed by a hub and its incident channels. Starfish rebalances among these incident channels to efficiently mitigate their balance depletion. 
In a nutshell, the major contributions of this paper are summarized as follows:

% In this study, we present Starfish, a rebalancing technique that improves rebalancing efficiency and channel capacity compared to existing approaches. Starfish is applicable to any node as it takes only the one-hop local topology of the node (the node is called a hub in the Starfish protocol to differentiate it from other nodes), a starfish-style structure centered at the hub,  and rebalances among the channels associated with the hub to overcome the balance depletion of any channel associated with the hub. Moreover, all the nodes in a PCN can run Starfish independently and simultaneously, as each node rebalances only the amounts it allocates to its associated channels. 
% In a nutshell, major contributions of this paper can be summarized as follows:
% % Starfish is applicable to any node as it takes only the one-hop local topology of the node (the node is called a hub in the Starfish protocol to differentiate it from other nodes), a starfish-style structure centered at the hub,  and rebalances among the channels associated with the hub to overcome the balance depletion of any channel associated with the hub. Moreover, all the nodes in a PCN can run Starfish independently and simultaneously, as each node rebalances only the amounts it allocates to its associated channels. 
% %Our Starfish protocol is specifically optimized for star-shaped network topologies, which often lead to the channel depletion issue. 
% %We have not only addressed this issue but also leveraged the high-degree nodes to merge multiple channels, achieving a performant rebalancing method. 

\begin{table*}[!htb]
	\begin{center}
	\begin{threeparttable}
		\caption{Comparison of Starfish with Other Approaches}
        \label{table:comparison}
        \tabcolsep=0.2cm
		\renewcommand{\arraystretch}{1.05}
        \setlength{\tabcolsep}{3.5pt}
        \begin{tabular}{l c c c c c c}
			\toprule[1pt]
	             &  \multicolumn{1}{c}{\textbf{\begin{tabular}[c]{@{}c@{}}\#TXs for \\ $n$-Time Refunding\end{tabular}}}
	             & \multicolumn{1}{c}{\textbf{\begin{tabular}[c]{@{}c@{}}Wait Time for \\ 1-Time Refunding\end{tabular}}}
             & \multicolumn{1}{c}{\textbf{\begin{tabular}[c]{@{}c@{}}Usability of \\ C/NC Cases\end{tabular}}}
             & \multicolumn{1}{c}{\textbf{\begin{tabular}[c]{@{}c@{}}\#OPTs for \\ $N$-Channel Rebalance\end{tabular}}}
             & \multicolumn{1}{c}{\textbf{\begin{tabular}[c]{@{}c@{}}Capacity \\ Bound\end{tabular}}}
             & \multicolumn{1}{c}{\textbf{\begin{tabular}[c]{@{}c@{}}Bitcoin- \\ compatible\end{tabular}}} \\
            \midrule[0.5pt]
            Revive~\cite{khalil2017revive} 
              & 0 & $\mathcal{O}(1)$ & \LEFTcircle & -- & $\mathcal{C}_{max}$ & Yes \\

	            Cycle~\cite{hong2022cycle} 
	              & $n$ & $\mathcal{O}(\Delta)$ & \LEFTcircle & -- & $\mathcal{C}_{max}$ & No \\
              
            Close-Open~\cite{poon2016bitcoin} 
              & $2n$ & $\mathcal{O}(\Delta)$ & \CIRCLE & -- & $\mathcal{C}_{max}$ & Yes \\

            LOOP~\cite{loop} 
              & $n$ & $\mathcal{O}(\Delta)$ & \CIRCLE & -- & $\mathcal{C}_{max}$ & Yes \\

            Shaduf~\cite{ge2022shaduf} 
              & 1 & $\mathcal{O}(1)$ & \CIRCLE & $N(N-1)^\dag$ & $\sum \mathcal{C}_{i}$ & No \\

            \textbf{Starfish} 
              & \textbf{1} & \bm{$\mathcal{O}(1)$} & \CIRCLE & \bm{$N$} & \bm{$\sum \mathcal{C}_{i}$} & No \\
			\bottomrule[1pt]
		\end{tabular}

		\begin{tablenotes}
		    \item[\#TX/\#OPT] \#TX denotes the number of on-chain transactions, while \#OPT denotes the number of on-chain interactions with channel contracts; $n$ denotes the number of refunding operations, while $N$ denotes the number of channels involved in rebalancing.
            \item[C/NC] Cycle-based / Non-cycle-based.
            \item[$\Delta$] The latency of transaction confirmation on a blockchain.
            \item[$\dag$] Shaduf's ``All-Bind'' strategy is required to enable full $N$-channel flexibility, incurring $N(N-1)$ OPTs; the ``All-to-One'' strategy needs $2(N-1)$ OPTs but does not support arbitrary pairwise rebalancing.
            \item[$\mathcal{C}_{max}$] The highest channel capacity.
		\end{tablenotes}
	\end{threeparttable}
	\end{center}
\end{table*}

\begin{itemize}
    \item \textbf{Efficient hub-centric rebalancing with linear on-chain cost.} 
    We design Starfish as a hub-centric rebalancing protocol that enables simultaneous rebalancing across multiple incident channels. 
    By introducing a novel merge contract, Starfish logically merges the hub’s adjacent channels into a single rebalancing pool, within which balances can be rebalanced arbitrarily and repeatedly in an off-chain manner after a one-time setup.
    This requires only $O(N)$ on-chain operations, where $N$ is the number of channels adjacent to the hub. 
    In contrast, Shaduf incurs $O(N^2)$ on-chain operations under pairwise binding. 
    % Consequently, Starfish substantially improves rebalancing efficiency in hub-dominated PCNs.
    
    % \item \textbf{High rebalancing efficiency.} Starfish offers an effective solution for rebalancing multiple channels associated with a node. We use a unique merge contract that requires only $N$ on-chain operations to connect several independent channels, resulting in a cost-effective rebalancing process. The traditional approach balances only two adjacent channels at a time. In contrast, our Starfish protocol can simultaneously merge and rebalance multiple channels based on their specific needs and channel connection conditions, thereby enhancing the efficiency of multi-channel rebalancing.
    % % Furthermore, it enables the largest rebalancing capacity possible by aggregating the total budget of all channels, surpassing the traditional approach of a fixed budget assigned to each channel.
    % \item \textbf{Large channel capacity.} Starfish improves channel scalability and efficiency by combining the capacities of multiple merged channels. This allows for the largest rebalancing capacity by utilizing the total budget of all channels. Additionally, Starfish promotes liquidity among merged channels, which is more effective than the traditional approach of assigning a fixed budget to each channel. These features enable the system to handle a significant volume of off-chain transactions with a high rate of success.
    \item \textbf{Formal security proof.} 
    We formalize the security properties of Starfish in the Universal Composability (UC) framework using the ideal/real world paradigm. 
    In particular, we define the required security guarantees for hub-centric rebalancing in the presence of malicious channel participants, and provide a formal UC security proof showing that Starfish satisfies these requirements.
    \item \textbf{Implementation and empirical evaluation.} 
    We implement a proof-of-concept prototype of Starfish on Ethereum and conduct comparative experiments against existing rebalancing protocols. 
    Our results show that when a hub is connected to $10$ channels, the on-chain gas cost of Shaduf’s binding operation is approximately $12.6\times$ that of Starfish’s merge operation. 
    Meanwhile, Starfish also achieves a higher payment success ratio than existing rebalancing protocols under representative workloads.
        
    % \item \textbf{Formal security proof and comparison study.} In order to validate the safety and liveness properties of Starfish when facing potential malicious channel participants, we formalize the protocol and use the universally composable framework to prove that Starfish satisfies all the necessary security requirements for rebalancing. By experiments, Starfish has shown superior performance in various typical conditions when compared to other similar protocols.
\end{itemize}

% \add{Question: Is our method Bitcoin compatible?}

\section{Related Work}

In a payment channel network, channel rebalancing refers to the process of redistributing balances across channels so as to ensure optimal liquidity and functionality. Table~\ref{table:comparison} presents a comprehensive comparative analysis on Starfish and other prevailing methods for rebalancing. 

% \textbf{Routing-based.} To alleviate channel imbalances, various routing strategies have been proposed. Spider~\cite{sivaraman2020high} uses a multi-path congestion control protocol, Fence~\cite{wang2023fence} and \cite{jiang2023balance} utilize a balance-aware fee-incentivized routing algorithm, \cite{liu2023balanced} constructs a comprehensive weight model using the Analytic Hierarchy Process, considering channel capacity, fees, and path length to alleviate channel congestion, and \cite{yang2023optimal} configures a dynamic routing algorithm for each PCH.

% \textbf{Rebalancing-based.} 
Existing rebalancing approaches can be broadly divided into cycle-based and non-cycle-based methods. 
Revive~\cite{khalil2017revive} performs rebalancing without on-chain transactions but relies on a cycle formed by payment channels: it sends a sequence of equal-amount off-chain payments along the cycle to shift balances away from depleted channels. A key drawback is that normal payments on the same cycle must be suspended during rebalancing. Cycle~\cite{hong2022cycle} alleviates this limitation by allowing rebalancing to proceed concurrently with regular payments. It distinguishes between local and global rebalancing states and employs a smart contract to record the global state and arbitrate conflicts; nevertheless, Cycle still incurs one on-chain transaction per refunding and requires ${O(\Delta)}$ consensus waiting time, where $\Delta$ denotes the blockchain confirmation latency. 
By contrast, Close-Open~\cite{poon2016bitcoin} rebalances a channel by reopening it through two on-chain transactions, whereas LOOP~\cite{loop} reduces this to a single on-chain transaction per rebalance. Consequently, their on-chain cost grows linearly with the number of refunding operations, and each operation must wait for ${O(\Delta)}$ confirmation latency. Importantly, neither approach relies on cycles, but they do not provide a mechanism to coordinate rebalancing across multiple incident channels at a hub. 

% Revive~\cite{khalil2017revive}'s rebalancing process does not require on-chain operations but relies on a cycle formed by payment channels. It sends a series of off-chain transactions with equal amounts in a cycle to rebalance channels that have been depleted. A drawback of Revive is that when rebalancing through transactions, normal transactions on the cycle should be stopped. Cycle~\cite{hong2022cycle} breaks this limitation, allowing the rebalancing process to be performed simultaneously with normal transactions. Cycle categorizes the rebalancing states into global and local ones, and relies on a smart contract to record the global state and arbitrate conflicts. For a $n$-time refunding, Cycle requires one on-chain operation and waits for ${O(\Delta)}$ on-chain consensus time. 
% Close-Open~\cite{poon2016bitcoin} rebalances a channel by reopening it through two on-chain operations, while LOOP~\cite{loop} reduces two to one. This implies that for a $n$-time refunding, $2n$ and $n$ on-chain transactions are respectively required. Note that these two approaches do not depend on cycles, and their on-chain consensus waiting time is ${O(\Delta)}$, where $\Delta$ is the latency of transaction confirmation on a blockchain. %Note that both Close-Open and LOOP do not depend on cycles.
% %
% %Similar to Revive, it depends on a circular topology. 

Shaduf~\cite{ge2022shaduf} enables multi-channel rebalancing without relying on any cycle. After a single on-chain Bind operation and ${O(\Delta)}$ confirmation latency, two bound channels can perform an arbitrary number of refunding operations off-chain using signatures alone. To apply Shaduf to $N$ channels, we consider three binding strategies: 
(1) ``High-to-Low''. Channels are paired according to their balances, yielding $N/2$ Bind/Unbind operations and thus $N$ on-chain operations.
(2) ``All-to-One''. All channels are bound to a pivot channel, requiring $N-1$ Bind/Unbind operations and $2(N-1)$ on-chain operations.
(3) ``All-Bind''. Every channel is bound to every other channel, requiring $N(N-1)/2$ Bind/Unbind operations and $N(N-1)$ on-chain operations.
These strategies provide different rebalancing capabilities. Although ``All-to-One'' has linear on-chain cost, it relies on a pivot channel and does not support direct rebalancing between two non-pivot channels. Therefore, it does not provide the same full $N$-channel rebalancing capability as Starfish and is not the basis of our on-chain overhead comparison. In contrast, ``All-Bind'' supports direct rebalancing between any pair of channels and provides the same full $N$-channel rebalancing capability as Starfish. Accordingly, our comparison is with ``All-Bind'': Shaduf requires $N(N-1)$ on-chain operations, whereas Starfish supports capacity reallocation between any two edges in the merge with only $N$ on-chain operations.

% Shaduf~\cite{ge2022shaduf}, through one on-chain binding operation, after waiting for ${O(\Delta)}$ on-chain consensus time, allows two channels to perform refunding $n$ times through off-chain signatures without relying on any cycle. For a mutual binding of multiple channels, Shaduf takes different binding strategies to determine the number of Bind/Unbind operations on-chain. Particularly, three typical binding strategies are adopted: (1)``High to Low'': users bind the channel with the highest balance to the one with the lowest balance, followed by the second-highest balance to the second-lowest balance, and so on, resulting in $N/2$ Bind/Unbind operations for $N$ channels and thus involving $N$ on-chain operations;
% (2) ``All to One'': all channels are bound to the same channel, resulting in $N-1$ Bind/Unbind operations and $2(N-1)$ on-chain operations;
% (3) ``All Bind'': each channel is bound to every other channel, resulting in $N(N-1)/2$ Bind/Unbind operations and $N(N-1)$ on-chain operations.
% Note that Close-Open, LOOP, Revive, and Cycle cannot rebalance multiple channels while Shaduf supports such a feature. With different binding strategies, each channel's capacity varies. However, the average capacity that can be used by each channel for any strategy cannot exceed $\mathcal{C}_{max}$.
% In comparison, Starfish consistently elevates the average transaction capacity per channel to $\sum \mathcal{C}_{i}$ and requires only $N$ on-chain operations.

\section{Preliminaries}
\label{sec:pre}
% \heading{Blockchain and Smart Contract.} Blockchain technology is a decentralized, distributed ledger that records transactions across a network of computers. Each block in the chain contains a timestamp and a cryptographic hash of the previous block, making it tamper-evident and providing an immutable record of all transactions. One of the most well-known blockchain platforms is Ethereum, which is a programmable blockchain that allows developers to build and deploy decentralized applications (DApps) and smart contracts. Smart contracts are self-executing contracts that run on the blockchain and automatically enforce the terms of the agreement. 

\textbf{Payment Channel. }A payment channel is a type of off-chain solution that allows two parties to establish a temporary channel for exchanging balances, conduct transactions within the channel, and record the final balances on the underlying blockchain. Such a channel does not require recording each transaction on the blockchain, and its lifecycle consists of three operations.

%A payment channel is a type of off-chain solution that allows two parties to conduct multiple transactions without recording each one on the underlying blockchain. Instead, the channel participants establish a temporary channel for exchanging assets, and then only the final balances are recorded on the blockchain. The lifecycle of a payment channel consists of three operations. 

\begin{packeditemize}
	\item \textbf{Open.} The process of opening a payment channel involves each party creating a smart contract that locks a certain amount of fund as its initial balance on the blockchain. 
 % These funds are then used to finance the payment channel. Each participant receives a private key from the channel to sign transactions. 

	\item \textbf{Update.} Once the channel is open, the participants can transact off-chain by signing transactions between themselves. 
 % Each transaction updates the balances of the channel and requires the signatures of both parties.

	\item \textbf{Close.} To close the channel, the participants must publish the final balances on the blockchain. This allows each party to receive the leftover of the fund that has been locked in the smart contract. 
 % The final balance can be either agreed upon by both parties or enforced by the smart contract's logic, depending on the design of the payment channel.

\end{packeditemize}

\heading{Payment Channel Networks.} A payment channel network is composed of interconnected payment channels, allowing two nodes in the network to complete transactions off-chain through a path formed by multiple adjacent channels. In such a network, nodes along a payment path can be categorized into two types: endpoint nodes and routing nodes. The two endpoint nodes act as the initiator and the receiver of a transaction, while routing nodes are responsible for facilitating transaction routing and may charge fees for this service. The network often exhibits a mesh topology, with many nodes having high degrees. Such nodes typically serve as routing nodes without actively initiating transactions. As a result, they are referred to as payment hubs. Payment hubs may frequently face channel depletion issues due to the presence of numerous skewed payments in a payment channel network.

%The network often exhibits a star-like topology, meaning there are many nodes with high degrees. Nodes with high degree typically serve as routing nodes without actively initiating transactions. As a result, they are also referred to as payment hubs. These payment hubs frequently face channel depletion issues due to the presence of numerous skewed payments in the payment channel network.

% \heading{Atomic Broadcast.} An atomic broadcast protocol involving $n$ nodes, denoted by $\mathsf{AtomicBroadcast}(\cdot)$, is designed to achieve the following properties with an overwhelming probability:
% \begin{packeditemize}
% 	\item \textbf{Agreement.} If any honest node receives a value $m$, then all honest nodes in the system will eventually receive the same value $m$.

% 	\item \textbf{Total Order.} For any two honest nodes respectively producing sequences of values $m_1, m_2, \ldots, m_j$ and $m'_1, m'_2, \ldots, m'_j$, $m_i=m'_i$ for all $1 \leq i \leq \min(j, j')$.

% 	\item \textbf{Validity.} If any honest node broadcasts a value $m$, then all honest nodes in the system will eventually receive it. 
%     % If a value $m$ is input to $n - f$ honest nodes, then all honest nodes eventually deliver $m$, where $f$ is the number of faulty nodes.
% \end{packeditemize}

\begin{figure*}[!htb]
\centering
\centerline{\includegraphics[width=0.95\linewidth]{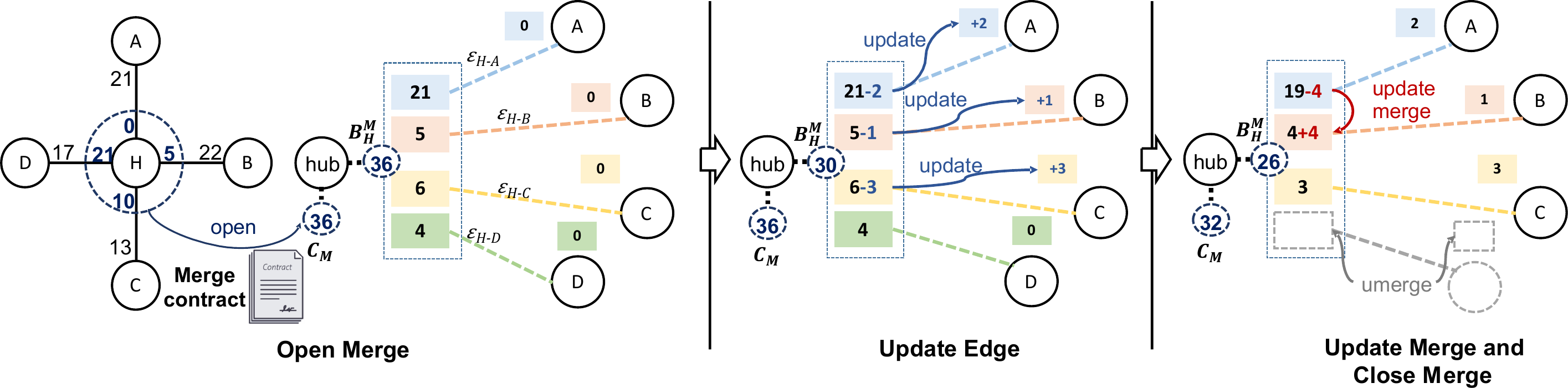}}
\caption{Procedures of Starfish, where $B_H^M$ and $C_M$ denote $H$'s total balance and the total capacity of the merge, respectively.}
\label{fig:starfish}
\end{figure*}

\section{The Design of Starfish}\label{sec:starfish}

%Starfish primarily involves four procedures: Open Merge, Update Edge, Update Edge, and Close Merge, along with three procedures related to channels: Open Channel, Update Channel and Close Channel.

% We demonstrate the design of Starfish protocol via a four-channel example illustrated in Fig.~\ref{fig:starfish}. Assume a node $H$ (also called hub in Starfish) has established four off-chain channels with users $A$, $B$, $C$, and $D$, denoted as $( H, A )$, $( H, B )$, $( H, C )$, and $( H, D )$, respectively. The node $H$ has deposited 0, 5, 10, and 21 units respectively in these channels. There exist two issues in this payment network: (1) $H$ is unable to transfer money to $A$ because $H$ has depleted its deposit in $(H, A)$; %This is the so-called channel depletion.
% (2) even though there is a total balance of 36 owned by $H$ in the payment network, $H$ cannot initiate a transaction of 36 because the balances in these channels are not shared. Starfish aims to rebalance these channels to allow the node $H$ to transact freely with any of the parties $A$, $B$, $C$, and $D$ using its total balance of 36. It primarily involves four procedures: Open Merge, Update Edge, Update Merge, and Close Merge, along with three procedures related to channels: Open Channel, Update Channel and Close Channel.

We illustrate Starfish using the four-channel example in Fig.~\ref{fig:starfish}. A hub $H$ maintains off-chain channels with users $A$, $B$, $C$, and $D$, denoted as $(H,A)$, $(H,B)$, $(H,C)$, and $(H,D)$, with $H$'s deposits being 0, 5, 10, and 21 units, respectively. This network exhibits two limitations: $H$ cannot pay $A$ due to channel depletion on $(H,A)$, and $H$ cannot leverage its total balance of 36 in a single transaction because balances are siloed across channels. Starfish rebalances these channels to enable $H$ to transact with any user using the aggregate balance, and consists of four core procedures: Open Merge, Update Edge, Update Merge, and Close Merge (in addition to standard channel operations).

\subsection{Open Merge: Creating Starfish from Scratch}

Open Merge merges the balances of all channels incident to a single node (i.e., the hub) into a unified funding pool, whose capacity is defined as the sum of the involved channel balances. This merge operates over a starfish-like structure (Fig.~\ref{fig:starfish}, Open Merge) and is implemented by a smart contract, denoted as $\texttt{MERGE}$. During an Open Merge involving $N$ channels, $\texttt{MERGE}$ interacts with each incident channel exactly once to extract the hub's balance. Although the Merge operation is invoked once, these interactions correspond to $N$ on-chain operations. After the merge, the collective capacity of the starfish structure remains constant throughout subsequent off-chain trading. For example, in Fig.~\ref{fig:starfish} (Open Merge), $H$ pledges its balances in $(H,A)$, $(H,B)$, $(H,C)$, and $(H,D)$ to $\texttt{MERGE}$, yielding $B_H^M=C_M=36$.
Enabling full rebalancing across $N$ channels via traditional approaches typically relies on the ``All-Bind'' strategy, under which every pair of channels must be bound to leverage the aggregate capacity. This incurs $N(N-1)$ on-chain operations, as each binding involves two channels. In contrast, Starfish touches each channel only once during Open Merge, achieving the same aggregate rebalancing capability with only $N$ on-chain operations.

After collecting all balances into $\texttt{MERGE}$, off-chain trading can proceed. The primary challenge is to prevent \textit{double spending}, especially when the hub colludes with end users. For instance, if $H$ concurrently transacts with $A$ and $D$ for amounts 18 and 19, their sum exceeds $H$'s total balance $B_H^M=36$ in the merge, constituting a double-spending attack. A direct defense is to enforce a global total order of off-chain transactions via consensus; however, this approach incurs substantial communication overhead and may render the hub, which must process transactions sequentially, a performance bottleneck.

% After collecting all balances to the $\texttt{MERGE}$ contract, off-chain trading can be enabled. Nevertheless, two prominent challenges commonly emerge in this context. %Herein, we delineate their potential manifestations within the Starfish protocol. 
% (1) Excessive payment: It is necessary to ensure that each newly issued transaction does not exceed the remaining balance of the payment account. Such a scenario can occur due to network delay or errors in balance counting. Each member of the group must confirm to this requirement. In this case, a smart contract should be able to determine which transaction to reject and finish final settlements safely. 
% %
% (2) Double spending: Starfish should avoid double-spending attacks, especially when the hub colludes with end-users. For example, if the hub $H$ transacts with $A$ and $D$ for amounts of 18 and 19, respectively, the sum is more than the total balance it possesses (i.e., 36), a double-spending attack occurs. A commonly used approach to counter such an attack is to apply consensus algorithms. However, if we adopt a consensus protocol to ensure the full order of all off-chain transactions, it would lead to high communication overhead and the hub that orderly deals with transactions can be a bottleneck. 

Starfish avoids global consensus by partitioning the pooled capacity into per-edge allocations during Open Merge. As shown in Fig.~\ref{fig:starfish} (Open Merge), the total capacity 36 is apportioned into 21, 5, 6, and 4 units and assigned to edges $\epsilon_{H\text{-}A}$, $\epsilon_{H\text{-}B}$, $\epsilon_{H\text{-}C}$, and $\epsilon_{H\text{-}D}$, respectively, enabling each hub--user pair to trade off-chain within its assigned allocation. These allocations can be guided by historical transaction volumes (e.g., heavier traffic between $H$ and $A$ warrants a larger allocation on $\epsilon_{H\text{-}A}$). Moreover, to support dispute resolution without a global ordering, each edge is associated with a monotone version number, denoted as $\mathsf{versionE}$, which will be used to order and validate off-chain updates on that edge.

% To solve these challenges, we partition the overall capacity of the funding pool into smaller allocations during the open merge procedure. As depicted in Fig.~\ref{fig:starfish} (Open Merge), the total capacity of 36 is apportioned into 21, 5, 6, and 4 units, with each being assigned to a specific edge, facilitating the connection between the hub and an individual end user. For instance, the capacity of 21 is allocated to the edge identified as $\epsilon_{H-A}$, thereby establishing the initial merge balances of $H$ and $A$ on $\epsilon_{H-A}$ as 21 and 0 units, respectively. The allocation of capacity to individual edges is mainly determined by the preceding transaction volumes. In cases where the transaction volume between entities $H$ and $A$ is substantial, the capacity of the $\epsilon_{H-A}$ edge should be considerable. After this capacity partitioning, each edge in the starfish structure is allocated with certain amount of balance (the result of rebalancing), namely the capacity of the edge,  with which the associated users can trade off-chain. Furthermore, to counter the double-spending attack, each edge is associated with a version number termed $\mathsf{versionE}$. 

\subsection{Update Edge: Treatment to Double Spending}

Update Edge is the off-chain procedure for updating the state on a designated edge by reallocating its merge balances between the hub and an end user, while conserving the edge capacity assigned in Open Merge. As illustrated in Fig.~\ref{fig:starfish} (Update Edge), $H$ transfers 2, 1, and 3 units to $A$, $B$, and $C$ on $\epsilon_{H-A}$, $\epsilon_{H-B}$, and $\epsilon_{H-C}$, respectively. After these updates, $B_H^M$ decreases from 36 to 30, while $C_M$ remains 36. Each update, jointly signed by the two endpoint parties, increments the edge-specific version number $\mathsf{versionE}$. The pairwise signatures together with the monotone $\mathsf{versionE}$ impose a total order of updates within each edge and serve as verifiable evidence for the smart contract to resolve disputes and finalize settlements, thereby preventing double spending without enforcing a global consensus across all edges and allowing updates to proceed in parallel.

\subsection{Update Merge and Close Merge: Rebalancing Capacity for Improved Availability}
% In the Starfish protocol, the facilitation of capacity transfers occurs seamlessly through an off-chain update merge operation that is straightforward to implement. This operation is initiated by a request, prompting two edges to transfer their capacities. Subsequently, this transfer is recorded in an off-chain transaction necessitating the signatures of the hub and the two involved end users. Upon broadcast, the transaction requires global consensus for approval. To ensure the sequential integrity of all update merge transactions, an atomic broadcast involving all users is employed. Upon successful completion of the atomic broadcast, indicating unanimous consensus, the update merge transaction is validated and assigned a version number denoted as $\mathsf{versionM}$. This methodology ensures the security of the update merge process while circumventing costly on-chain operations. Given the expediency of atomic broadcasts and the infrequent occurrences of update merge transactions compared to standard off-chain transactions, the associated overhead remains acceptable. For instance, 
% as depicted in Fig.~\ref{fig:starfish} (Update Merge and Close Merge), $\epsilon_{H-A}$ transfer the capacity of 4 to $\epsilon_{H-B}$ by a update merge operation. 

In Starfish, capacity can be rebalanced across edges via an off-chain update merge operation. An update merge transfers capacity between two edges and is recorded as an off-chain transaction jointly signed by the hub and the two involved end users. The hub then runs an atomic broadcast to synchronize all users in the merge to the new merge state. The remaining end users do not sign the update; they receive the atomic broadcast, verify the state, and may challenge inconsistencies. Upon successful completion of the atomic broadcast, the transaction is validated and assigned a monotone version number $\mathsf{versionM}$. Since update merge operations are infrequent compared to standard edge updates, the overhead of atomic broadcast remains acceptable. For example, in Fig.~\ref{fig:starfish} (Update Merge and Close Merge), $\epsilon_{H-A}$ transfers 4 units of capacity to $\epsilon_{H-B}$ via an update merge. This operation reallocates capacity between the two edges, while $B_H^M$ and $C_M$ remain 30 and 36, respectively.

To handle user departures and safely dismantle merge structures, Starfish provides a close merge operation that unmerges affected edges. During close merge, users verify the latest edge and merge states using $\mathsf{versionE}$ and $\mathsf{versionM}$ and can challenge inconsistencies, enabling correct dispute resolution and settlement. In Fig.~\ref{fig:starfish} (Update Merge and Close Merge), $D$ exits the merge, reducing $C_M$ from 36 to 32 and $B_H^M$ from 30 to 26. The 4 units removed from the merge are returned to $H$'s balance in channel $(H,D)$. After close merge, users retain ownership of their underlying channels; any user may subsequently close a channel on-chain, and channels participating in a merge must first undergo close merge before termination.

% Additionally, the close merge operation safeguards against the potential disruption of the nodes or the dissolution of merge structures upon the departure of end users. It involves disentangling merged channels and is initiated by a request for unmerging, requiring the involvement of the end-users of the affected edges. Throughout the close merge procedure, users within the edge verify the update edge and merge states by inspecting $\mathsf{versionE}$ and $\mathsf{versionM}$, addressing any inconsistencies through challenges. The combined utilization of $\mathsf{versionE}$ and $\mathsf{versionM}$ ensures the full ordering of transactions. In Fig.~\ref{fig:starfish} (Update Merge and Close Merge), $D$ exits the merge, decreasing the total capacity from 36 to 32.  
% %
% The close merged users retain ownership of the off-chain channels. Furthermore, initiating the closure of channels is within the purview of any user and involves the termination of a channel. The closure process can present complexities, especially when channels are integrated into a merge structure, requiring prior close merge procedure. 

\section{Formal Modeling of Starfish}
\label{security_analysis}

This section outlines the formal model used to analyze the Starfish protocol, covering the network assumptions, security objectives, and the specific notations for channels, merges, and edges. 
\subsection{Model}
The network has a fixed set of parties $\mathcal{P}=\{P_1,...,P_n\}$, where all parties are rational non-myopic players \cite{tsabary2021mad}. 
% We assume that the setup phase of the Public-Key Infrastructure (PKI) is conducted by a trusted party prior to the initiation of the protocol. 
A message $(m,\sigma_P)$ is deemed valid if $\sigma_P$ constitutes a valid signature of party $P$ on $m$. Similar to Perun \cite{dziembowski2019perun}, we consider a synchronous network where messages are exchanged in a round-based manner. Each message between parties is delivered within one round. For instance, a message sent by $A$ in round $r$ can be received by $B$ before the beginning of round $r+1$. Communications between the parties and the environment $\mathcal{E}$ as well as the ideal functionality $\mathcal{F}$ are instantaneous, with messages sent through secure channels that prevent tampering. The adversary $\mathcal{A}$ can see the messages from honest parties but cannot alter them. 
We model a static adversary $\mathcal{A}$ that corrupts arbitrary parties before the protocol begins. Corruption means that $\mathcal{A}$ gets all internal states of the corrupted parties and takes full control of them. We denote the transaction confirmation delay as $\Delta$. We use $\tau$ to represent the time points related to on-chain contract operations, and $t$ to denote the time points of off-chain operations.

\subsection{Security Goals}
Our primary security goal is to ensure the protection of the balances of honest parties, guaranteeing that their funds remain uncompromised. More specifically, our protocol requires a unanimous consent from both parties when opening and updating channels; a collective agreement from all involved parties during the open merge; mutual consent from the parties involved in the update edge; signatures from the hub and the two involved end users during the update merge, followed by an atomic broadcast to synchronize all users in the merge; the processes of closing merge and channel should be completed within a reasonable time, and it is imperative to ensure that the coin shift is conducted based on the latest state of the honest parties. 
The detailed properties are delineated below: 

\begin{itemize}
\item \textbf{Consensus on Open Channel and Update Channel: }Within a channel, both opening and updating require mutual agreements from the involved parties. The opening of a channel can be completed within $O(\Delta)$ time, whereas updating the channel’s state is achieved in a constant time.

\item \textbf{Consensus on Open Merge and Update Edge:} The open merge process involves multiple channels associated with an intermediate party, requiring the consent of all parties involved. The open merge process requires $O(\Delta)$ time to complete. For the update edge, both parties must confirm each merge state, which takes constant time to complete.

\item \textbf{Consensus on Update Merge:} The update merge process entails the transfer of capacity between two edges and requires signatures from the hub and the two involved end users. Subsequently, the hub runs an atomic broadcast to synchronize the update merge state among all users. The remaining end users may challenge inconsistencies but are not required to sign the update. The entire update merge operation should take constant time.

\item \textbf{Guaranteed Close Merge and Close Channel:} A party associated with an edge can request to close merge and conduct a coin shift for the channel. The close merge process takes $O(\Delta)$ time. Any party in the channel can request to close the channel, which also takes $O(\Delta)$ time.

\item \textbf{Guaranteed balance payout for users:} The process to close merge and close a channel must settle accounts according to the latest state involving the honest party.
\end{itemize}

\subsection{Notations}
This section defines the core components and operations within our payment channel network model. We describe the properties of individual channels, merge operations that involve multiple connections, and the edges within those merges.

\subsubsection{Payment Channel (\texorpdfstring{$\beta$}{beta})}
A fundamental element is the payment channel, denoted by $\beta$, which facilitates transactions directly between two parties. Each channel $\beta$ is characterized by:

\begin{itemize}
    \item \text{Unique identifier:} $\beta.id \in \{0,1\}^*$. A unique string that distinguishes this channel from all others in the network.
    \item \text{Participating parties:} $\beta.\mathsf{users} = (A, B)$. The ordered pair identifying the two parties authorized to transact using this channel.
    \item \text{Channel balances:} $\beta.\mathsf{balanceC} : \beta.\mathsf{users} \rightarrow \mathbb{R}^{\ge0}$. This function maps each party ($A$ or $B$) to their current non-negative balance within the channel. The total funds remain constant within the channel during internal payments.
    \item \text{Channel Payment ($\theta$):} A payment operation within the channel is represented by a function $\theta:\beta.\mathsf{users}\rightarrow\mathbb{R}$. This function specifies the amount transferred, satisfying the conservation property $\theta(A)+\theta(B)=0$. Applying the payment $\theta$ updates the balances such that the new balance for each party $X \in \{A, B\}$ becomes $\beta.\mathsf{balanceC}(X) + \theta(X)$. We denote this update collectively as $\beta.\mathsf{balanceC} \leftarrow \beta.\mathsf{balanceC} + \theta$.
    % \item \textbf{Merge Set:} $\beta.\mathsf{mergeSet}$ - This set stores references to all merges involving the channel.
    \item \text{Channel version number:} $\beta.\mathsf{versionC}$. An integer that increments with each update to the channel's state (e.g., after a payment $\theta$), used for state synchronization and conflict resolution.

    \item \text{Merge set:} $\beta.\mathsf{mergeSet}$. This set contains all the merge contracts associated with the channel $\beta$. 
    % \item \textbf{State Signatures:} $\beta.\mathsf{sig}$ - This collection of signatures from both parties signifies their agreement on the current channel state.
\end{itemize}

\subsubsection{Merge (\texorpdfstring{$\varphi$}{phi})}
A merge, denoted by $\varphi$, represents a higher-level construct, potentially involving multiple users connected through a central point or hub via individual edges. A merge is defined by:

\begin{itemize}
    \item \text{Unique identifier:} $\varphi.\tilde{id} \in \{0,1\}^*$. A unique string identifying this specific merge instance.
    \item \text{Merged Users:} $\varphi.\mathsf{users}$. The set containing all users participating in this merge, typically connected to a common hub.
    \item \text{Merged edges:} $\varphi.\mathsf{edges} = (\epsilon_1, \dots, \epsilon_n)$. A tuple listing all the individual edge connections (\textit{defined below}) that constitute this merge contract.
    \item \text{Merge version number:} $\varphi.\mathsf{versionM}$. An integer that increments when the merge state itself is updated (e.g., through a re-merge or capacity reallocation), tracking the evolution of the merged structure.
    % \item \textbf{State Signatures:} $\varphi.\mathsf{sig}$ - This collection of signatures from involved parties signifies their agreement on the current merge state.
\end{itemize}

\subsubsection{Edge within a Merge (\texorpdfstring{$\epsilon$}{epsilon})}
Each edge $\epsilon$ within a merge's edge list ($\epsilon \in \varphi.\mathsf{edges}$) represents a specific bilateral relationship, usually between a hub and one user within the merge context. An edge has the following attributes:

\begin{itemize}
    \item \text{Participating users:} $\epsilon.\mathsf{users} = (\mathsf{hub}, \mathsf{user})$. An ordered pair identifying the two endpoints of this edge, typically the central hub and a specific user from $\varphi.\mathsf{users}$.
    \item \text{Edge capacity:} $\epsilon.\mathsf{capacity}\in\mathbb{R}^{\ge0}$. A non-negative value representing the total capacity locked into this specific edge connection.
    \item \text{Edge balances:} $\epsilon.\mathsf{balanceE}: \epsilon.\mathsf{users} \rightarrow \mathbb{R}^{\ge0}$. This function maps the hub and the user to their respective non-negative balances within this edge. The sum of these balances must always equal the edge's capacity: $\epsilon.\mathsf{balanceE}(\mathsf{hub}) + \epsilon.\mathsf{balanceE}(\mathsf{user}) = \epsilon.\mathsf{capacity}$.
    \item \text{Edge Payment ($\tilde{\theta}$):} A payment along the edge is defined by a function $\tilde{\theta}:\epsilon.\mathsf{users}\rightarrow\mathbb{R}$, satisfying $\tilde{\theta}(\mathsf{hub})+\tilde{\theta}(\mathsf{user})=0$. Applying $\tilde{\theta}$ updates the edge balances: $\epsilon.\mathsf{balanceE} \leftarrow \epsilon.\mathsf{balanceE}+\tilde{\theta}$.
    \item \text{Edge version number:} $\epsilon.\mathsf{versionE}$. An integer that increments after each update specific to this edge (e.g., after an edge payment $\tilde{\theta}$), tracking the state progression of the individual edge.
\end{itemize}

\subsubsection{Merge Update (\texorpdfstring{$\hat{\theta}$}{theta-hat})}
An update merge operation, denoted $\hat{\theta}$, facilitates the reallocation of capacity between two edges, say $\epsilon_1$ and $\epsilon_2$, that are part of the same merge $\varphi$. This operation is defined as a function: $\hat{\theta}: \varphi.\mathsf{edges} \;\rightarrow\; \mathbb{R}$. 
This function specifies the change in capacity for edges within the merge. For a reallocation between $\epsilon_1$ and $\epsilon_2$, it must satisfy:
$
\hat{\theta}(\epsilon_1) \;+\; \hat{\theta}(\epsilon_2) \;=\; 0
$
and $\hat{\theta}(\epsilon_j) = 0$ for all other edges $\epsilon_j \in \varphi.\mathsf{edges} \setminus \{\epsilon_1, \epsilon_2\}$. Applying $\hat{\theta}$ updates the capacities of the involved edges as follows:
$
\epsilon_i.\mathsf{capacity} \leftarrow \epsilon_i.\mathsf{capacity} + \hat{\theta}(\epsilon_i), \quad \text{for } i \in \{1,2\}.
$
For brevity, we may denote this collective capacity update for the affected edges $\epsilon \in \{\epsilon_1, \epsilon_2\}$ simply as $\epsilon.\mathsf{capacity}+\hat{\theta}$.

\section{Detailed Starfish Protocol}
This section provides a detailed description of the Starfish protocol's implementation, including the functioning of the channel and merge smart contracts, the definition of the ideal functionality for security analysis, and the step-by-step execution of the real-world protocol.

\subsection{Channel Contract and Merge Contract}

We have used the UC security model \cite{canetti2001universally} to demonstrate the security of our protocol. This model defines two worlds: the ideal world and the real world. In the real world, parties execute the protocol $\Pi$ while facing an adversary $\mathcal{A}$ and interacting with the contract functionality $\mathcal{C}$. In the ideal world, the idealized protocol, known as ideal functionality $\mathcal{F}$, is executed through interactions with parties and a simulator $\mathcal{S}$, which simulates the behaviors of the adversary. All parties receive inputs from and send outputs to the environment $\mathcal{E}$. The protocol $\Pi$ is considered UC-secure if the environment $\mathcal{E}$ cannot computationally distinguish whether it is interacting with the protocol in the real world or the one in the ideal world. We define $\lambda$ as the security parameter, and $\mathsf{EXEC}^{\Pi,\mathcal{C}}_{\mathcal{E},\mathcal{A}}$ ($\lambda$) denotes the output of environment $\mathcal{E}$ executing the real world protocol $\Pi$ with adversary $\mathcal{A}$ in the $\mathcal{C}$-hybrid world. $\mathsf{IDEAL}^{\mathcal{F}}_{\mathcal{E},\mathcal{S}}(\lambda)$ denotes the output of environment $\mathcal{E}$ executing the ideal functionality $\mathcal{F}$ with simulator $\mathcal{S}$ in the ideal world. The formal security definition is as follows:
\begin{definition}
Protocol $\Pi$ executing in the $\mathcal{C}$-hybrid world UC-realizes the ideal functionality $\mathcal{F}$ with respect to the global ledger $\mathcal{L}$ and with blockchain delay $\Delta$, if for any PPT adversary $\mathcal{A}$ there exists a simulator $\mathcal{S}$ such that 
\begin{equation}
\mathsf{EXEC}^{\Pi,\mathcal{C}}_{\mathcal{E},\mathcal{A}}(\lambda)\approx\mathsf{IDEAL}^{\mathcal{F}}_{\mathcal{E},\mathcal{S}}(\lambda)
\end{equation}
where $\approx$ denotes the computational indistinguishability.
\end{definition}

\textit{Ledger and contract functionalities:} The ledger functionality $\mathcal{L}$ is designed as a foundational, transparent, and immutable record of the balances $x_i$ for each party $P_i$ in the system. The core design principle is to establish a single, publicly observable source of truth for asset ownership. A key design choice is the indirect manipulation of balances: parties cannot directly alter the ledger. Instead, all updates (adding or removing coins) are exclusively triggered by the contract functionality $\mathcal{C}$ (in the real world) or its ideal counterpart $\mathcal{F}$ (in the ideal world). This ensures that all ledger modifications adhere to the predefined logic embedded within smart contracts, enhancing security and control. The ledger itself maintains a simple state as a tuple of balances $(x_1, ..., x_n)$ and supports two basic operations: $\mathtt{add}$ (always successful for non-negative amounts) and $\mathtt{remove}$ (only successful if sufficient balance exists), thereby preventing negative balances at the base layer. The ledger operates passively, responding only to commands from other functionalities. 

\myonecolumnbox{Ledger Functionality  $\mathcal{L}$}{
        \textbf{Initialization: }The ledger functionality is initialized by a message $(x_1,...,x_n) \in \mathbb{R}_n^{\ge 0}$ from the environment $\mathcal{E}$, where $x_i$ denotes the coins of party $P_i$ in $\mathcal{L}$; the tuple is stored in the ledger.

        \textbf{Adding coins: }Upon receiving a message $(\mathtt{add}, P_i, y)$ (where $P_i \in \mathcal{P}$ and $y \in \mathbb{R}^{\ge 0}$), the ledger functionality updates $x_i := x_i + y$. 

        \textbf{Removing coins: }Upon receiving a message $(\mathtt{remove}, P_i, y)$ (where $P_i \in \mathcal{P}$ and $y \in \mathbb{R}^{\ge 0}$), the ledger functionality updates $x_i := x_i - y$ if $x_i \ge y$; otherwise, ignores the message and stops.
}{Ledger Functionality \label{ledger:function}}

The essence of $\mathcal{C}$'s design lies in its role as a neutral intermediary ensuring the consistent state transitions of channels and merges by interacting with the underlying ledger $\mathcal{L}$ and the involved parties. For channel establishment, it enforces mutual agreement within a defined timeframe. For merges, it facilitates the linking and unlinking of separate contracts, adjusting balances accordingly. During channel closure, it guarantees that both parties agree on the final state before the funds are released back to them, providing a secure and reliable mechanism for managing these collaborative agreements.

\mylongbox{Contract Functionality $\mathcal{C}$}{
    \begin{center}
        (A) \textbf{The contract execution of channel $\beta$}
    \end{center}
    Wait for the following messages: 
    \begin{enumerate}
        \item Upon receiving $(\mathtt{open},\beta)$ from $A$ in $\tau$, send $(\mathtt{remove},A,\beta.\mathsf{balanceC}(A))$ to ledger $\mathcal{L}$ and send $(\mathtt{opening},\beta)$ to $B$ in $\tau$. Wait for one of the following messages:
        
        \begin{enumerate}
            \item Upon receiving $(\mathtt{open},\beta)$ from $B$ within $\tau_1\le\tau+\Delta$, send $(\mathtt{remove},B,\beta.\mathsf{balanceC}(B))$ to $\mathcal{L}$ in $\tau_1$, output $(\mathtt{opened})$ to $\beta.\mathsf{users}$.
            \item Otherwise, send $(\mathtt{add},A,\beta.\mathsf{balanceC}(A))$ to $\mathcal{L}$ after $\tau_2>\tau+\Delta$ and close the contract. Output $(\mathtt{not\text{-}opened})$ to $A$. 
            
        \end{enumerate}
        
        \item Upon receiving $(\mathtt{chan\text{-}merge},\varphi)$ from $\mathcal{C}(\varphi.\tilde{id})$ in $\tau$, add $\varphi$ to $\beta.\mathsf{mergeSet}$, set $\beta.\mathsf{balanceC}(\mathsf{hub})$ $:=$ $\beta.\mathsf{balanceC}(\mathsf{hub})$ $-$ $\epsilon.\mathsf{capacity}$ and stop. 
        
        \item Upon receiving $(\mathtt{chan\text{-}closeM},\varphi,\mathsf{msgE})$ from $\mathcal{C}(\varphi.\tilde{id})$ in $\tau$, remove $\varphi$ from $\beta.\mathsf{mergeSet}$. Set $\beta.\mathsf{balanceC}$ $:=$ $ \beta.\mathsf{balanceC}+ \epsilon.\mathsf{balanceE}$, and stop. 
        
        \item Upon receiving ($\mathtt{closeC}$, $id$, $\mathsf{msgC}$) from $A$ in $\tau$, if $\mathsf{msgC}$ is valid, store $\mathsf{msgC}$. Send ($\mathtt{closeC}$, $id$) to $B$. 
        Upon receiving ($\mathtt{closeC}$, $id$, $\mathsf{msgC}$) from $B$ within $\tau_1$ $\le$ $\tau$ $+4\Delta$, if $\mathsf{msgC}$ of $B$ is valid and $\beta^{(B)}.\mathsf{versionC}$ $>$ $\beta^{(A)}.\mathsf{versionC}$, store $B$'s $\mathsf{msgC}$ and discard the old one.        
        Send $(\mathtt{add},A,\beta.\mathsf{balanceC}(A))$ and ($\mathtt{add}$, $B$, $\beta.\mathsf{balanceC}(B)$) to $\mathcal{L}$, output ($\mathtt{closedC}$) to $\beta.\mathsf{users}$.
    
    \end{enumerate}
    
    \begin{center}
        (B) \textbf{The contract execution of merge $\varphi$}
    \end{center}

    Wait for the following messages:
    \begin{enumerate}               
        \item Upon receiving $(\mathtt{merge},\varphi,\Sigma)$ from $\mathsf{hub}$ in $\tau$, let $\beta$ be the underlying channel for each edge $\epsilon\in\varphi.\mathsf{edges}$. If $\beta.\mathsf{balanceC}(\mathsf{hub})$ $\ge$ $\epsilon.\mathsf{capacity}$, set $\epsilon.\mathsf{balanceE}(\mathsf{hub})$ $:=$ $\epsilon.\mathsf{capacity}$. Send $(\mathtt{chan\text{-}merge},\varphi)$ to $\mathcal{C}(\beta.id)$ in $\tau$, output $(\mathtt{merged})$ to all users and stop. Otherwise, output $(\mathtt{not\text{-}merged})$ to all users and stop. 
        
        \item Upon receiving ($\mathtt{closeM}$, $\tilde{id}$, $\epsilon$, $\mathsf{msgM}$, $\mathsf{msgE}$) from $\mathsf{hub}$ in $\tau$, if $\mathsf{msgM}$ and $\mathsf{msgE}$ are valid states, store them, and send ($\mathtt{closingM}$, $\tilde{id}$, $\epsilon$) to $\mathsf{user}$. Wait for the following messages: 
        
        \begin{enumerate}
            \item Upon receiving ($\mathtt{closeM}$, $\tilde{id}$, $\epsilon$, $\mathsf{msgM}$, $\mathsf{msgE}$) from $\mathsf{user}$ in $\tau_1\le\tau+\Delta$, if $\mathsf{msgM}$ and $\mathsf{msgE}$ of $\mathsf{user}$ are valid, $\varphi^{(\mathsf{user})}.\mathsf{versionM}$ $>$ $\varphi^{(\mathsf{hub})}.\mathsf{versionM}$ and $\epsilon^{(\mathsf{user})}.\mathsf{versionE}$ $>$ $\epsilon^{(\mathsf{hub})}.\mathsf{versionE}$, store $\mathsf{user}$'s $\mathsf{msgM}$ and $\mathsf{msgE}$ and discard old ones. Send ($\mathtt{closeM\text{-}check}$, $\tilde{id}$, $\mathsf{msgM}$) to $\mathsf{users}$. 
            
            \item Upon receiving $(\mathtt{timeout},\tilde{id})$ after $\tau_2>\tau+\Delta$ from $\mathsf{hub}$, send ($\mathtt{closeM\text{-}check}$, $\tilde{id}$, $\mathsf{msgM}$) to $\mathsf{users}$. 
            
            \item Upon receiving ($\mathtt{closeM\text{-}challenge}$, $\tilde{id}$, $\mathsf{msgM}$) from $R$ within $\tau_3\le\tau+2\Delta$, if $R$'s $\mathsf{msgM}$ is a valid state and $\varphi^{(R)}.\mathsf{versionM}$ is the highest, keep $R$'s $\mathsf{msgM}$, discard the old one, set $\epsilon.\mathsf{balanceE}$ of $\mathsf{msgE}$ to be equal to $\epsilon.\mathsf{capacity}$ of $\mathsf{msgM}$. 
        \end{enumerate}            
        Remove $\mathsf{user}$ from $\varphi.\mathsf{users}$. Remove $\epsilon$ from $\varphi.\mathsf{edges}$.  Send $(\mathtt{chan\text{-}closeM},\varphi,\mathsf{msgE})$ to $\mathcal{C}(\beta.id)$. Output ($\mathtt{closedM}$, $\tilde{id}$, $\epsilon$) to all users and stop.
    \end{enumerate}

    %  \begin{center}
    %     (C) \textbf{Subroutine for closing channel $\beta$}
    % \end{center}

    % Send $(\mathtt{add},P,\beta.\mathsf{balance}(P))$ and $(\mathtt{add},Q,\beta.\mathsf{balance}(Q))$ to ledger $\mathcal{L}$, output $(\mathsf{closed})$ to $\beta.\mathsf{users}$ and close the channel. 
}{Contract functionality $\mathcal{C}$: (A) The contract execution of channel $\beta$; (B) The contract execution of merge $\varphi$.}{fig:contract:channel}

\subsection{Ideal Functionality and Real Protocol}

The ideal functionality $\mathcal{F}$ communicates with the party set $\mathcal{P}$, the simulator $\mathcal{S}$, and the ledger $\mathcal{L}$. To simplify the protocol description without compromising its security, we impose certain restrictions on the inputs provided by the environment $\mathcal{E}$. Specifically, the environment $\mathcal{E}$ should not provide illogical inputs, such as requesting the activation of a channel that is already open or requesting a channel balance that exceeds the user's actual coins in the ledger. 

Our ideal functionality $\mathcal{F}$ comprises seven procedures. (A) \textit{Open Channel} activates a channel only after both parties agree. (B) \textit{Update Channel} allows either party to request a state update, which is completed only with the consent of the other party. (C) \textit{Open Merge} is initiated by the hub and completes only after all end users agree. (D) \textit{Update Edge} transfers balances within an edge subject to approval by the parties on that edge. (E) \textit{Update Merge} shifts merged capacity between two edges, requires signatures from the hub and the two involved end users, and then runs an atomic broadcast so that all users can synchronize the merge state and challenge inconsistencies. (F) \textit{Close Merge} can be initiated by either party on an edge and returns the edge capacity to the underlying channel. (G) \textit{Close Channel} can be initiated by either channel party and ultimately transfers the channel balances back to the ledger $\mathcal{L}$.
The complete specification of the ideal functionality and the corresponding security analysis are presented in Appendix~\ref{app:security-proof}.

% Next, we provide an analysis regarding how the ideal functionality $\mathcal{F}$ achieves the security properties. 

% Then we provide a comprehensive description on the real-world Starfish protocol as depicted in Fig.~\ref{Procedure 1} and Fig.~\ref{Procedure 2}.

% Then we describe the real-world Starfish protocol, as summarized in Fig.~\ref{Procedure 1} and Fig.~\ref{Procedure 2}. 
% We omit the standard Open Channel and Update Channel procedures, since they follow conventional payment-channel designs; our focus is on the merge-related operations that are specific to Starfish.

\quad \newline
\mylongboxTwoColumn{Starfish Protocol}{   
    Assume $A$ is the caller and $B$ is the responder. Let $\beta^{(A)}$ denote the channel state form $A$'s perspective, and similarly for $B$. 
    
    \begin{center}
        (A) \textbf{Open Channel}
    \end{center}\label{open channel}

    \begin{enumerate}[label=(\arabic*)]
        \item \textbf{[A]} Upon receiving $(\uline{\mathtt{open}},\beta)$ from the environment $\mathcal{E}$ in $t$, $A$ initializes a new channel contract instance $\mathcal{C}(\beta.id)$ and sends a $\Delta\text{-}\mathsf{bounded}$ contract construction message $(\mathtt{open},\beta)$, then proceeds to step (4).  

        \item \textbf{[B]} Upon receiving $(\mathtt{opening,\beta})$ from $\mathcal{C}(\beta.id)$ within $\tau\le t+\Delta$, $B$ sends $(\uline{\mathtt{opening}},\beta)$ to $\mathcal{E}$. $B$ sends $(\mathtt{open},\beta)$ to $\mathcal{C}(\beta.id)$ after receiving (\uline{$\mathtt{open}$}) from $\mathcal{E}$ in $\tau$, then proceeds to step (3).

        \item \textbf{[B]} Upon receiving $(\mathtt{opened})$ from $\mathcal{C}(\beta.id)$ within $\tau_1\le \tau+\Delta$, $B$ outputs (\uline{$\mathtt{opened}$}) and stops. 

        \item \textbf{[A]} Upon receiving $(\mathtt{opened})$ from $\mathcal{C}(\beta.id)$ within $\tau_2\le t+2\Delta$, $A$ outputs (\uline{$\mathtt{opened}$}) and stops. If receiving ($\mathtt{not\text{-}opened}$) from $\mathcal{C}(\beta.id)$ after $\tau_3>t+2\Delta$, $A$ outputs (\uline{$\mathtt{not\text{-}opened}$}) and stops.
        
    \end{enumerate}

    \begin{center}
        (B) \textbf{Update Channel}
    \end{center}

    \begin{enumerate}[label=(\arabic*)]
        \item \textbf{[A]} Upon receiving $(\uline{\mathtt{updateC}},id,\theta)$ from $\mathcal{E}$ in $t$, $A$ sets $\mathsf{msgC}=(\beta^{(A)}.\mathsf{versionC}+1,\beta^{(A)}.\mathsf{balanceC}+\theta)$ and signs it with $\sigma_A$. $A$ sends $(\mathtt{updateC},id,\mathsf{msgC},\sigma_A)$ to $B$ and proceeds to step (3). 
        
        \item \textbf{[B]} Upon receiving $(\mathtt{updateC},id,\mathsf{msgC}=(v,b),\sigma_A)$ from $A$ in $t_1:=t+1$, $B$ checks $v\overset{?}{=}\beta^{(B)}.\mathsf{versionC}+1$. If the condition holds, $B$ sends $(\uline{\mathtt{updateC\text{-}req}},id,b-\beta^{(B)}.\mathsf{balanceC})$ to $\mathcal{E}$. Upon receiving $(\uline{\mathtt{updateC\text{-}ok}})$ from $\mathcal{E}$, $B$ generates the signature $\sigma_B$ on $\mathsf{msgC}$ and sends $(\mathtt{updateC},\sigma_{B})$ to $A$ in $t_1$. If the condition does not hold, $B$ ignores the message.

        \item \textbf{[A]} Upon receiving $(\mathtt{updateC},\sigma_{B})$ from $B$ in $t_2:=t_1+1$, $A$ outputs $(\uline{\mathtt{updatedC}})$ and stops; otherwise, $A$ outputs $(\uline{\mathtt{not\text{-}updatedC}})$ and stops.

    \end{enumerate}

    \begin{center}
        (G) \textbf{Close Channel}
    \end{center}

    \begin{enumerate}[label=(\arabic*)]
        \item \textbf{[A]} Upon receiving $(\uline{\mathtt{closeC}},id)$ from $\mathcal{E}$ in $t$, if $\beta.\mathsf{mergeSet}\neq\emptyset$, $A$ executes procedure (F) for each $\varphi\in \beta.\mathsf{mergeSet}$. After closing all merges, $A$ sets $\mathsf{msgC}= $ ($\beta^{(A)}.\mathsf{versionC}$, $\beta^{(A)}.\mathsf{balanceC}$, $\Sigma_C$), where $\Sigma_C$ is the signature set of the current channel state. $A$ sends ($\mathtt{closeC}$, $id$, $\mathsf{msgC}$) to $\mathcal{C}(\beta.id)$ and goes to step (3). 
                        
        \item \textbf{[B]} Upon receiving $(\mathtt{closingC},id)$ from $\mathcal{C}(\beta.id)$ in $\tau$, if $\beta.\mathsf{mergeSet}\neq\emptyset$, $B$ executes procedure (F) for each $\varphi\in \beta.\mathsf{mergeSet}$. After closing all merges, $B$ generates $\mathsf{msgC}$ like $A$ and sends ($\mathtt{closeC}$, $id$, $\mathsf{msgC}$) to $\mathcal{C}(\beta.id)$ and goes to step (3).

        \item \textbf{[A, B]} Upon receiving $(\mathtt{closedC})$ from $\mathcal{C}(\beta.id)$ within $\tau_1\le\tau+4\Delta$, $A$ and $B$ output $(\uline{\mathtt{closedC}})$ and stops. 
        
    \end{enumerate}

}{Channel Operations \label{Procedure 1}}

\quad \newline
\mylongboxTwoColumn{Starfish Protocol}{   

    \begin{center}
        (C) \textbf{Open Merge}
    \end{center}

    % Let $\mathsf{users}$ denote $\varphi.\mathsf{users}$, and $P\in\mathsf{users}$. 

   \begin{enumerate}[label=(\arabic*)]
       \item \textbf{[Hub]} Upon receiving $(\uline{\mathtt{merge}},\varphi)$ from $\mathcal{E}$ in $t$, $\mathsf{hub}$ generates $\sigma_{\mathsf{hub}}$ for $(\varphi,t)$, sends $(\mathtt{merge},\varphi,\sigma_{\mathsf{hub}})$ to $P\in\varphi.\mathsf{users}$ in $t$, and goes to step (3). 
       
       \item \textbf{[Users]} Upon receiving $(\mathtt{merge},\varphi,\sigma_{\mathsf{hub}})$ from $\mathsf{hub}$ in $t_1:=t+1$, $P$ sends $(\uline{\mathtt{merge\text{-}req}},\varphi)$ to $\mathcal{E}$. Upon receiving (\uline{$\mathtt{merge\text{-}ok}$}) from $\mathcal{E}$, $P$ generates $\sigma_P$ for $(\varphi,t)$, sends $(\mathtt{merge},\varphi,\sigma_P)$ to $\mathsf{hub}$, and goes to step (5). 
                      
       \item \textbf{[Hub]} Upon receiving all $(\mathtt{merge},\varphi,\sigma_P)$ in $t_2:=t_1+1$, $\mathsf{hub}$ sends $(\uline{\mathtt{merge\text{-}confirm}},\varphi,\Sigma)$ to $\mathcal{E}$, where $\Sigma$ collects signatures from all users. 
       Upon receiving $(\uline{\mathtt{merge\text{-}confirmed}})$ from $\mathcal{E}$, $\mathsf{hub}$ creates a merge contract $\mathcal{C}(\varphi.\tilde{id})$ and sends a $\Delta\text{-}\mathsf{bounded}$ message $(\mathtt{merge},\varphi,\Sigma)$ to $\mathcal{C}(\varphi.\tilde{id})$. Then $\mathsf{hub}$ goes to step (4). 
       
       \item \textbf{[Hub]} Upon receiving $(\mathtt{merged}, \varphi)$ from $\mathcal{C}(\varphi.\tilde{id})$ within $\tau\le t_2+\Delta$, $\mathsf{hub}$ updates $\mathsf{balanceC}$ and $\mathsf{balanceE}$ according to the contracts. $\mathsf{hub}$ outputs $(\uline{\mathtt{merged}}, \varphi)$ and stops. Otherwise, $\mathsf{hub}$ outputs $(\uline{\mathtt{not\text{-}merged}})$ and stops.
       
       \item \textbf{[Users]} Upon receiving $(\mathtt{merged}, \varphi)$ from $\mathcal{C}(\varphi.\tilde{id})$ within round $\tau\le t_1+\Delta+1$, $P$ updates $\mathsf{balanceC}$ and $\mathsf{balanceE}$ according to the contracts. $P$ outputs $(\uline{\mathtt{merged}})$, then stops; otherwise, $P$ outputs $(\uline{\mathtt{not\text{-}merged}})$ after receiving $(\mathtt{not\text{-}merged})$ from $\mathcal{C}(\varphi.\tilde{id})$, and stops.

   \end{enumerate}

    \begin{center}
        (D) \textbf{Update Edge}
    \end{center}

    \begin{enumerate}[label=(\arabic*)]
        \item \textbf{[Hub]} Upon receiving (\uline{$\mathtt{updateE}$}, $\tilde{id}$, $\tilde{\theta}$, $\epsilon$) from $\mathcal{E}$ in $t$, $\mathsf{hub}$ sets $\mathsf{msgE}=(\epsilon^{(\mathsf{hub})}.\mathsf{versionE}+1,\epsilon^{(\mathsf{hub})}.\mathsf{balance
        E}+\tilde{\theta})$ and signs it with $\sigma_{\mathsf{hub}}$. $\mathsf{hub}$ sends ($\mathtt{updateE}$, $\tilde{id}$, $\mathsf{msgE}$, $\sigma_{\mathsf{hub}}$) to $\mathsf{user}$ and proceeds to step (3). 
                        
        \item \textbf{[User]} Upon receiving ($\mathtt{updateE}$, $\tilde{id}$, $\mathsf{msgE}=$($v$, $b$), $\sigma_{\mathsf{hub}}$) from $\mathsf{hub}$ in $t_1:=t+1$, $\mathsf{user}$ checks $v$ $\overset{?}{=}$ $\epsilon^{(\mathsf{user})}.\mathsf{versionE}$$+1$. If the condition holds, $\mathsf{user}$ sends (\uline{$\mathtt{updateE\text{-}req}$}, $\tilde{id}$, $b-\epsilon^{(\mathsf{user})}.\mathsf{balanceE}$) to $\mathcal{E}$. $\mathsf{user}$ updates $\epsilon^{(\mathsf{user})}.\mathsf{versionE}:=v$ and $\epsilon^{(\mathsf{user})}.\mathsf{balanceE}:=b$ after receiving (\uline{$\mathtt{updateE\text{-}ok}$}) from $\mathcal{E}$. $\mathsf{user}$ generates $\sigma_{\mathsf{user}}$ on $\mathsf{msgE}$ and sends ($\mathtt{updateE}$, $\sigma_{\mathsf{user}}$) to $\mathsf{hub}$, and stops. Otherwise, $\mathsf{user}$ ignores the message.

        \item \textbf{[Hub]} Upon receiving ($\mathtt{updateE}$, $\sigma_{\mathsf{user}}$) from $\mathsf{user}$ in $t_2:=t_1+1$, $\mathsf{hub}$ outputs (\uline{$\mathtt{updatedE}$}), and stops; otherwise, $\mathsf{hub}$ outputs (\uline{$\mathtt{not\text{-}updateE}$}).

    \end{enumerate}  
    
    \begin{center}
        (E) \textbf{Update Merge}
    \end{center}

    Let $\epsilon_1$ and $\epsilon_2$ be the edges involved, connecting to end-users $P$ and $Q$, respectively. Let $\mathsf{cap}$ denote the set of all edge capacities, and let $R\in\varphi.\mathsf{users}$. 
    
    \begin{enumerate}[label=(\arabic*)]
        \item \textbf{[Hub]} Upon receiving (\uline{$\mathtt{updateM}$}, $\tilde{id}$, $\hat{\theta}$, $\epsilon_1$, $\epsilon_2$) from $\mathcal{E}$ in $t$, $\mathsf{hub}$ sets $\mathsf{msgM}=$ ($\varphi^{(\mathsf{hub})}.\mathsf{versionM}+1$, $\mathsf{cap}^{(\mathsf{hub})}$) and $\mathsf{msgE}=$ ($\epsilon_1^{(\mathsf{hub})}.\mathsf{versionE}+1$, $\epsilon_2^{(\mathsf{hub})}.\mathsf{versionE}+1$, $\epsilon^{(\mathsf{hub})}.\mathsf{capacity}+\hat{\theta}$, $\epsilon^{(\mathsf{hub})}.\mathsf{balanceE}(\mathsf{hub})+\hat{\theta}$), and generates $\sigma_{\mathsf{hub}}$ on them. $\mathsf{hub}$ sends $(\mathtt{updateM},\tilde{id},\mathsf{msgM},\mathsf{msgE},\sigma_{\mathsf{hub}})$ to $P$, $Q$, and goes to step (3).

        \item \textbf{[Users]} Upon receiving ($\mathtt{updateM}$, $\tilde{id}$, $\mathsf{msgM}=$($vM$, $c$), $\mathsf{msgE}=$($vE_1$, $vE_2$, $\hat{c}$, $\hat{b}$), $\sigma_{\mathsf{hub}}$) in $t_1:=t+1$, $P$ (or $Q$) checks $vM$$\overset{?}{=}$$\varphi^{(P)}.\mathsf{versionM}$$+1$, $vE_1$$\overset{?}{=}$$\epsilon_1^{(P)}.\mathsf{versionE}$$+1$. 
        If the condition holds, $P$ sends (\uline{$\mathtt{updateM\text{-}req}$}, $\tilde{id}$, $\hat{c}-\epsilon^{(P)}.\mathsf{capacity}$, $\epsilon_1$, $\epsilon_2$) to $\mathcal{E}$. 
        Upon receiving (\uline{$\mathtt{updateM\text{-}ok}$}) from $\mathcal{E}$, $P$ generates $\sigma_P$ for $\mathsf{msgM}$ and $\mathsf{msgE}$, sends $(\mathtt{updateM},\sigma_P)$ to $\mathsf{hub}$, and goes to step (4). 
        Otherwise, $P$ ignores the message. 

        \item \textbf{[Hub]} Upon receiving $(\mathtt{updateM},\sigma_{P(Q)})$ from both $P$ and $Q$ in $t_2:=t_1+1$, $\mathsf{hub}$ sends (\uline{$\mathtt{updateM\text{-}confirm}$}) to $\mathcal{E}$. 
        Upon receiving (\uline{$\mathtt{updateM\text{-}confirmed}$}) from $\mathcal{E}$, $\mathsf{hub}$ runs $\mathsf{AtomicBroadcast}(\mathtt{updateM},\tilde{id},\mathsf{msgM},\mathsf{msgE},\Sigma)$ to $\varphi.\mathsf{users}$, where $\Sigma$ collects the above signatures. If the $\mathsf{AtomicBroadcast}$ succeeds, $\mathsf{hub}$ outputs (\uline{$\mathtt{updatedM}$}) in $t_3:=t_2+2$ and stops; otherwise, $\mathsf{hub}$ outputs (\uline{$\mathtt{not\text{-}updatedM}$}), and stops.

        \item \textbf{[Users]} Upon receiving ($\mathtt{updateM}$, $\tilde{id}$, $\mathsf{msgM=}$($vM$, $c$), $\mathsf{msgE=}$($vE_1$, $vE_2$, $\hat{c}$, $\hat{b}$), $\Sigma$), $R$ checks $vM\overset{?}{=}\varphi^{(R)}.\mathsf{versionM}+1$. If the condition holds, $R$ broadcasts the message; if not, $R$ sends (\uline{$\mathtt{updateM\text{-}pending}$}) to $\mathcal{E}$. Upon receiving (\uline{$\mathtt{updateM\text{-}wrong}$}) from $\mathcal{E}$, $R$ ignores the message. If the $\mathsf{AtomicBroadcast}$ succeeds, $R$ updates both the $\mathsf{capacity}$ of each edge and $\mathsf{versionM}$ to remain consistent with $\mathsf{hub}$. If $R$ is either $P$ or $Q$, $R$ then updates $\mathsf{versionE}$ and $\mathsf{balanceE}$ of the corresponding edge to remain consistent with $\mathsf{hub}$. $R$ outputs (\uline{$\mathtt{updatedM}$}) in $t_3$ and stops; otherwise, $R$ outputs (\uline{$\mathtt{not\text{-}updatedM}$}), and stops.

        % \item $\epsilon_1^{(\mathsf{hub})}.\mathsf{capacity}+\hat{\theta}(\epsilon_1)$ and $\epsilon_2^{(\mathsf{hub})}.\mathsf{capacity}-\hat{\theta}(\epsilon_2)$

        % \item $\varphi^{(\mathsf{hub})}.\mathsf{capacity}(\epsilon_1)+\hat{\theta}(\epsilon_1)$ and $\varphi^{(\mathsf{hub})}.\mathsf{capacity}(\epsilon_1)-\hat{\theta}(\epsilon_1)$

        % \item $\epsilon_1^{(\mathsf{hub})}.\mathsf{balanceE}(\mathsf{hub})+\hat{\theta}(\epsilon_1)$ and $\epsilon_2^{(\mathsf{hub})}.\mathsf{balanceE}(\mathsf{hub})-\hat{\theta}(\epsilon_2)$

    \end{enumerate}

    \begin{center}
        (F) \textbf{Close Merge}
    \end{center}

    Assume $\mathsf{hub}$ is the caller and $\mathsf{user}$ is the responder. Let $\mathsf{cap}$ denote the set of all edge capacities. Let $\mathsf{users}$ denote $\varphi.\mathsf{users}\backslash\{\mathsf{user}\}$, and $R\in\mathsf{users}$. 

    \begin{enumerate}[label=(\arabic*)]
        \item \textbf{[Hub]} Upon receiving $(\uline{\mathtt{closeM}},\tilde{id},\epsilon)$ from $\mathcal{E}$ in $t$, $\mathsf{hub}$ sets $\mathsf{msgM}=$ ($\varphi^{(\mathsf{hub})}.\mathsf{versionM}$, $\mathsf{cap}^{(\mathsf{hub})}$, $\Sigma_R$), where $\Sigma_R$ is the signature set of the current merge state. $\mathsf{hub}$ sets $\mathsf{msgE}=$ ($\epsilon^{(\mathsf{hub})}.\mathsf{versionE}$, $\epsilon^{(\mathsf{hub})}.\mathsf{balanceE}$, $\Sigma_E$), where $\Sigma_E$ is the signature set of the current edge state. $\mathsf{hub}$ sends ($\mathtt{closeM}$, $\tilde{id}$, $\epsilon$, $\mathsf{msgM}$, $\mathsf{msgE}$) to $\mathcal{C}(\varphi.\tilde{id})$ in $t$, then goes to step (3).

        \item \textbf{[User]} Upon receiving $(\mathtt{closingM},\tilde{id},\epsilon)$ from $\mathcal{C}(\varphi.\tilde{id})$ in $\tau$, $\mathsf{user}$ generates $\mathsf{msgM}$ and $\mathsf{msgE}$ like $\mathsf{hub}$. 
        $\mathsf{user}$ sends ($\mathtt{closeM}$, $\tilde{id}$, $\epsilon$, $\mathsf{msgM}$, $\mathsf{msgE}$) to $\mathcal{C}(\varphi.\tilde{id})$ in $\tau$, and goes to step (5). 

        \item \textbf{[Hub]} If not receiving $(\mathtt{closedM},\tilde{id},\epsilon)$ within $2\Delta$ time, $\mathsf{hub}$ sends $(\mathtt{timeout},\tilde{id})$ to $\mathcal{C}(\varphi.\tilde{id})$, and goes to step (5). 

        \item \textbf{[Users]} Upon receiving ($\mathtt{closeM\text{-}check}$, $\tilde{id}$, $\mathsf{msgM}=(v,c)$) from $\mathcal{C}(\varphi.\tilde{id})$ in $\tau_1\le\tau+\Delta$, $R$ checks if $\varphi^{(R)}.\mathsf{versionM}>v$. If the condition holds, $R$ generates $\mathsf{msgM}$ like $\mathsf{hub}$. 
        $R$ sends ($\mathtt{closeM\text{-}challenge}$, $\tilde{id}$, $\mathsf{msgM}$) to $\mathcal{C}(\varphi.\tilde{id})$ and goes to step (6); otherwise, $R$ goes to step (6). 

        \item \textbf{[Hub, User]} Upon receiving ($\mathtt{closedM}$, $\tilde{id}$, $\epsilon$) from $\mathcal{C}(\varphi.\tilde{id})$ within $\tau_2\le\tau+2\Delta$, $\mathsf{hub}$ and $\mathsf{user}$ remove $\varphi$, update $\mathsf{balanceC}$ based on the contracts, output $(\uline{\mathtt{closedM}},\tilde{id},\epsilon)$ and then stop.

        \item \textbf{[Users]} Upon receiving ($\mathtt{closedM}$, $\tilde{id}$, $\epsilon$) from $\mathcal{C}(\varphi.\tilde{id})$ in $\tau_2$, $R$ updates $\varphi$ based on the contracts. $R$ outputs $(\uline{\mathtt{closedM}},\tilde{id},\epsilon)$ and then stops. 
                        
    \end{enumerate}

}{Merge Operations \label{Procedure 2}}

Then we describe the real-world Starfish protocol, as summarized in Fig.~\ref{Procedure 1} and Fig.~\ref{Procedure 2}. 
Initially, we outline the procedure for opening a channel, as shown in Procedure (A), with the corresponding contract functionalities illustrated in Fig.~\ref{fig:contract:channel}. Upon receiving $(\uline{\mathtt{open}},\beta)$ from the environment $\mathcal{E}$, $A$ sends a request to the contract instance $\mathcal{C}(\beta.id)$ for channel activation. Subsequently, $\mathcal{C}(\beta.id)$ informs $B$ about the initiation of the channel. If $B$ agrees to open the channel within $\Delta$ time, $\mathcal{C}(\beta.id)$ records the opening of channel $\beta$, reallocates the channel balances from ledger $\mathcal{L}$ into the contract, and outputs $(\mathtt{opened})$ to both $A$ and $B$. Otherwise, $\mathcal{C}(\beta.id)$ outputs $(\mathtt{not\text{-}opened})$ to $A$.

The process of channel updating is depicted in Procedure (B). Updating the channel does not require interaction with the contract. Upon receiving $(\uline{\mathtt{updateC}},id,\theta)$ from $\mathcal{E}$, $A$ first increments its local version number (i.e., $\beta.\mathsf{versionC}$), and updates the latest balance allocation based on $\theta$. $A$ then sends the updated state along with its signature to $B$, requesting a channel state update. After verifying the correctness of the signature and the state version number, $B$ seeks approval from $\mathcal{E}$ for the channel update. If $\mathcal{E}$ approves the update, $B$ responds to $A$ with an updated state signed by its signature. If $B$ does not respond, $A$ outputs $(\uline{\mathtt{not\text{-}updatedC}})$ and terminates the update channel process.

The procedure for open merge is shown in Procedure (C), with the corresponding contract functionalities detailed in Fig.~\ref{fig:contract:channel}. The open merge process requires initiation by the hub. To prevent replay attacks, the hub first signs the open merge message $(\varphi,t)$ and sends this signed open merge request to the end users. Upon receiving the request, each end user verifies the validity of the signature and then requests permission to merge from $\mathcal{E}$. Once permission is granted, each end user forwards the merge message along with its own signature back to the hub. The hub aggregates all the signatures and sends the complete merge message, now containing the collective signatures, to the contract $\mathcal{C}(\varphi.\tilde{id})$. Subsequently, $\mathcal{C}(\varphi.\tilde{id})$ invokes the relevant channel instances to record the merge information and outputs $(\mathtt{merged})$ to all users.

Following the merge process, parties can transact using the edge balances, as shown in Procedure (D). Upon receiving $(\uline{\mathtt{updateE}},\tilde{id},\tilde{\theta})$ from $\mathcal{E}$, the hub first increments its local edge version number $\epsilon.\mathsf{versionE}$ by 1 and adjusts the edge balance allocation according to $\tilde{\theta}$. The hub then sends the updated edge state, along with its signature, to the corresponding end user. After verifying the correctness of the updated merge state number and signature, the end user requests confirmation from $\mathcal{E}$ for the state update. Once the confirmation is received, the end user replies to the hub with a signed update edge message. If the end user does not respond, the hub outputs $(\uline{\mathtt{not\text{-}updatedE}})$ and terminates the process.

The update merge process, depicted in Procedure (E), transfers capacity between two edges and is initiated by the hub. Upon receiving $(\uline{\mathtt{updateM}}, \tilde{id}, \hat{\theta}, \epsilon_1, \epsilon_2)$ from $\mathcal{E}$, the hub increments the merge version $\varphi.\mathsf{versionM}$, updates the capacity allocation according to $\hat{\theta}$, and sends the proposed merge state to the end users of $\epsilon_1$ and $\epsilon_2$ (denoted $P$ and $Q$) for approval and signatures. The hub then runs an atomic broadcast within the merge group to synchronize all users to the new merge state: if the broadcast succeeds, all users accept the update, refresh their local $\varphi.\mathsf{versionM}$ and the capacities of $\epsilon_1,\epsilon_2$, and the parties on these edges additionally update the corresponding edge states (e.g., $\epsilon.\mathsf{versionE}$ and $\epsilon.\mathsf{balanceE}$); otherwise, the update is rejected and outputs $(\uline{\mathtt{not\text{-}updatedM}})$.

The close merge procedure, shown in Procedure (F), removes an edge from the merge contract and is implemented by the merge-contract instance $\mathcal{C}(\varphi.\tilde{id})$ (see Fig.~\ref{fig:contract:channel}). Upon receiving a close-merge request (e.g., from the hub), $\mathcal{C}(\varphi.\tilde{id})$ notifies the corresponding end user and waits up to $\Delta$ time: if the user responds with a newer valid close-merge state (with higher local edge/merge version numbers), the contract records it; otherwise, it defaults to the hub-provided state. The contract then queries the remaining participants for challenges and adopts the highest valid merge version number. Finally, letting $\beta$ denote the underlying channel corresponding to the edge $\epsilon$, $\mathcal{C}(\varphi.\tilde{id})$ invokes $\mathcal{C}(\beta.id)$ to reallocate channel balances according to the finalized highest valid edge and merge version numbers.

The procedure for close channel is illustrated in Procedure (G), with the related contract functions detailed in Fig. \ref{fig:contract:channel}. Upon receiving a close request from $\mathcal{E}$, $A$ submits it to $\mathcal{C}(\beta.id)$. If $\beta$ is still in a merge, $\mathcal{C}(\beta.id)$ first invokes Procedure (F) to close-merge the channel; otherwise, it notifies $B$ and waits up to $\Delta$ time for a response. The contract then settles according to the highest valid channel version number agreed by both parties if $B$ responds, or according to $A$'s latest valid state upon timeout, refunds the balances, and closes the channel.

% Keep the wide protocol box and its caption before the theorem and evaluation.
\begin{theorem}
\label{thm:uc-security}
The protocol Starfish executing in the $\mathcal{C}$-hybrid world UC-realizes the ideal functionality $\mathcal{F}$ with respect to the global ledger $\mathcal{L}$ and blockchain delay $\Delta$. 
\end{theorem}

\begin{proof}
    The complete simulation-based proof is provided in Appendix~\ref{app:security-proof}.
\end{proof}

\section{Evaluation}\label{impletation}

To demonstrate the practicality of Starfish, we implement a functional prototype on Ethereum, focusing on the feasibility and cost of the required on-chain operations.
Separately, to evaluate the impact of Starfish on payment channel networks, we conduct a simulation study using the Lightning Network dataset, which provides a large-scale, real-world PCN topology commonly adopted for evaluating rebalancing protocols, and compare Starfish against representative baselines, including Revive and Shaduf.

% To demonstrate the practicality of Starfish, we implemented a functional prototype leveraging the Ethereum platform. We then conducted a simulation study on the Lightning Network dataset to evaluate its performance within a payment channel network, contrasting our findings with those reported for Revive and Shaduf.

\subsection{Implementation} 
Our Starfish implementation is built on the Ethereum platform and comprises two contracts, namely the merge contract and the channel contract. The Starfish, Shaduf, and Revive projects have about 320, 440, and 260 LOCs, respectively. To use the Starfish protocol, users need to deploy only one merge contract, which serves them permanently, and consumes minimal on-chain resources. However, when creating new channels or making functional calls while running the protocol, there will be on-chain expenses. These expenses are calculated in gas, which is a unit of measurement used for quantifying the computational cost of transactions and smart contract executions on the Ethereum network. The initiator of a transaction needs to provide enough gas fee\footnote{The gas fee is calculated by multiplying the gas price, measured in Gwei, by the amount of gas required for the transaction or smart contract execution.}. Deploying the merge and channel contracts requires 1.8 million and 1.4 million gas, respectively.

\subsection{Gas Cost}

\subsubsection{Settings}
We vary the number of channels involved in rebalancing from 2 to 10, measuring the total gas cost of on-chain operations for Starfish, Shaduf and Revive. The implementation of Starfish requires Open/Close channels and Merge/Close merge contracts while that of Shaduf involves on-chain operations such as Open/Close channels and Bind/Unbind; but Revive only deals with Open/Close channel operations. Therefore, in our context, we compare Starfish with Shaduf and Revive, but won't include Revive's figures to save space.

In Starfish, we measure the gas costs of opening $N$ channels, merging all $N$ edges, unmerging all $N$ edges, and closing the $N$ channels. The $N$ edges require only one Merge operation and one Close Merge operation. Each of these operations interacts with all $N$ channel contracts and therefore corresponds to $N$ on-chain operations.
In Shaduf, the binding strategy determines the number of on-chain operations required by Bind and Unbind. Therefore, we explore three different binding strategies: ``High-to-Low'', ``All-to-One'', and ``All-Bind''. We respectively denote the Shaduf corresponding to these three strategies as HL-Shaduf, AO-Shaduf, and AB-Shaduf.
% , and estimate the overall gas cost for opening, binding, unbinding, and closing all channels according to each strategy.
In Revive, a globally impartial node oversees the rebalancing process. Only when there is a dispute over the rebalancing outcome while closing channels, on-chain operations related to rebalancing are needed. Therefore, we estimate the possible maximum consumption of gas in Revive when opening and closing $N$ channels.

\subsubsection{Results and Discussion} 

We analyze the assessment outcomes from two different angles.

\begin{figure*}[!htb]
\centering
\subfigure[Starfish]{
% \begin{minipage}[t]{0.25\linewidth}
\includegraphics[width=0.24\textwidth]{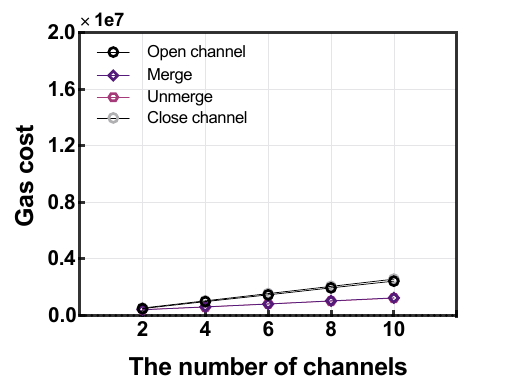}
\label{Starfish Gas Cost}
% \end{minipage}%
}%
\subfigure[HL-Shaduf]{
% \begin{minipage}[t]{0.25\linewidth}
\includegraphics[width=0.24\textwidth]{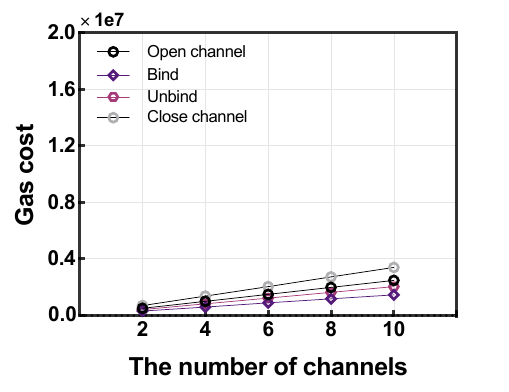}
\label{Shaduf HL Gas Cost}
% \end{minipage}%
}%
\subfigure[AO-Shaduf]{
% \begin{minipage}[t]{0.25\linewidth}
\includegraphics[width=0.24\textwidth]{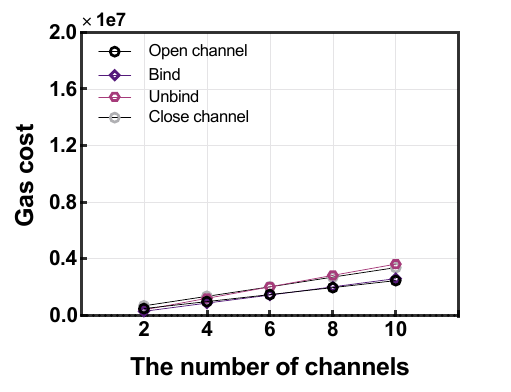}
\label{Shaduf AO Gas Cost}
% \end{minipage}
}%
\subfigure[AB-Shaduf]{
% \begin{minipage}[t]{0.25\linewidth}
\includegraphics[width=0.24\textwidth]{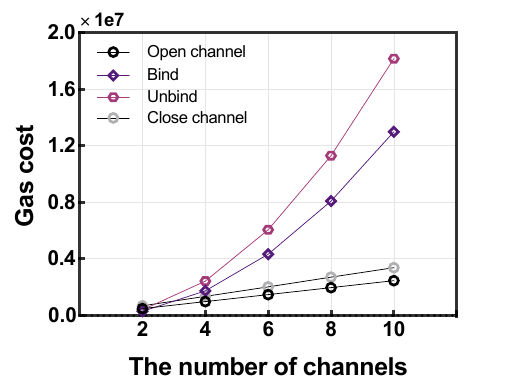}
\label{Shaduf AB Gas Cost}
% \end{minipage}
}%
\centering
\caption{Gas costs of Starfish and Shaduf under the High-to-Low (HL-Shaduf), All-to-One (AO-Shaduf), and All-Bind (AB-Shaduf) strategies, measured in gas.}
\end{figure*}

\textbf{Gas Cost of Starfish.} 
There are four types of on-chain operations for Starfish: Open Channel, Merge, Close Merge, and Close Channel. Fig.~\ref{Starfish Gas Cost} illustrates that the gas cost for opening and closing all channels is directly proportional to the number of channels. This is because the gas cost for a single Open Channel or Close Channel operation remains unchanged, but the number of channels impacts the total gas cost. 
The gas costs of merging and unmerging all $N$ edges also increase linearly with $N$. Although the $N$ edges require only one Merge operation and one Close Merge operation, each additional channel requires storing the state of the new channel and performing an additional signature verification within these operations. Consequently, the gas cost increases at a constant rate with the addition of one channel.

\label{fig:gas cost comparison}

\textbf{Cost Comparison.} 
Among the three Shaduf strategies, AB-Shaduf supports direct rebalancing between any pair of channels and therefore provides the most direct comparison with Starfish. HL-Shaduf and AO-Shaduf are also reported because they require fewer bindings but do not support direct rebalancing between every pair of channels.
The reported gas-cost ratios include all Bind and Unbind operations required by each Shaduf strategy and one Merge and one Close Merge operation for Starfish; Open Channel and Close Channel operations are shown separately and are not included in these ratios.
We first compare Shaduf's Bind/Unbind operations with Starfish's Merge/Close Merge operations. The gas cost for binding and unbinding all channels scales linearly with the number of channels in HL-Shaduf (Fig.~\ref{Shaduf HL Gas Cost}) and AO-Shaduf (Fig.~\ref{Shaduf AO Gas Cost}), while in AB-Shaduf (Fig.~\ref{Shaduf AB Gas Cost}), it follows a quadratic relationship with the number of channels. This is because the gas cost for Bind and Unbind operations remains constant, while noting that for HL-Shaduf, AO-Shaduf, and AB-Shaduf, $N$ channels respectively require $N$, $2(N-1)$, and $N(N-1)$ on-chain operations.
It's important to highlight that, in HL-Shaduf and AO-Shaduf, the slope of the two lines is respectively $1.6\times$ and $3.3\times$ that of Starfish. This is because the linear growth in Shaduf is due to multiple Bind/Unbind operations, which result in higher gas cost compared to Starfish, because of the additional storage and computation required.
When the number of channels reaches 10, the gas cost for Bind/Unbind operations in HL-Shaduf, AO-Shaduf, and AB-Shaduf is approximately $1.4\times$, $2.5\times$, and $12.6\times$, respectively, compared to Starfish's Merge/Close Merge. 
% In terms of Open channel and Close channel operations, both Revive and Shaduf show a constant increase in gas cost with the addition of one channel, which indicates a linear growth concerning the number of channels. 
% The slopes of these two lines for Revive and Shaduf are $1.2\times$ and $1.1\times$ that of Starfish, respectively.

\subsection{Success Ratio}\label{sec:success ratio}

\subsubsection{Settings} 
To enhance simulation realism, we use a snapshot of the real-world Lightning Network topology collected on 15 October 2025, containing approximately 12,632 active nodes and 43,758 payment channels. For all methods, since the initial balance allocation within each channel is invisible, we evenly divide each channel capacity between its two users. The channel capacities recorded in the snapshot form the $1\times$ baseline, and $k\times$ means that each channel capacity is multiplied by $k$. Under uniform sampling, payment initiators and receivers are sampled uniformly; under skewed sampling, initiators are selected according to an exponential distribution, while receivers are selected uniformly at random, and a higher payment skewness factor indicates more skewed payments. Payment values are randomly sampled from Bitcoin transaction data from September 1 to September 30, 2025. Values below the median channel capacity at $1\times$ are defined as small payments, while values equal to or above it are defined as large payments, resulting in 2.65 million and 6.65 million samples, respectively. Each payment follows a shortest path, and rebalancing is triggered when a channel on the path has insufficient balance. For Revive, the channel with insufficient balance is the target channel; it is rebalanced using a source channel of the same node by selecting the shortest cycle containing both channels. For Shaduf, each node binds its channels according to the corresponding binding strategy. In HL-Shaduf, the channel with the highest balance is bound to the channel with the lowest balance, followed by the second-highest and the second-lowest, and so on. In AO-Shaduf, all other channels are bound to the channel with the highest balance. In AB-Shaduf, every pair of channels is bound. When a channel is bound to multiple channels, the balance assigned to these bindings is evenly distributed among them. For Starfish, the node's balances in all its incident channels are merged. We evaluate HL-Shaduf, AO-Shaduf, AB-Shaduf, and Starfish using the same set of 50,000 payments, repeat each experiment ten times, and report the average success ratio.

\subsubsection{Results} We investigate how the success rates of six off-chain payment methods, with LN (Lightning Network) as a baseline, are affected by their respective capacity, skewness, and payment size.

% Keep explanatory text between the two full-width evaluation figures.
\begin{figure*}[!t]
\centering
\subfigure[Uniform small payments]{
\includegraphics[width=0.24\textwidth]{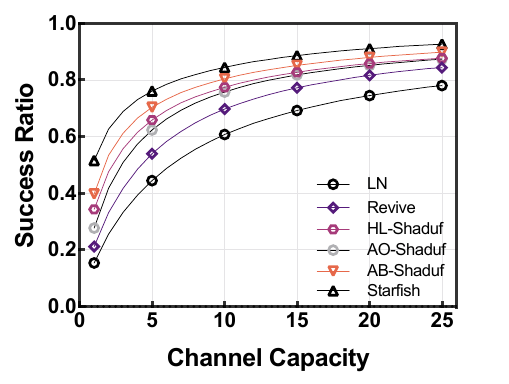}
\label{fig:Starfish Uniform Small Payment}
}%
\subfigure[Skew small payments]{
\includegraphics[width=0.24\textwidth]{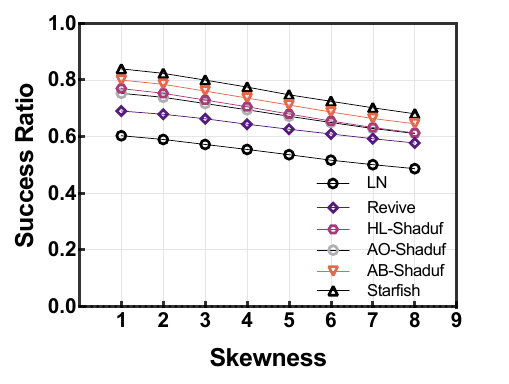}
\label{fig:Starfish Skew Small Payment}
}%
\subfigure[Fixed Skewness for small payments]{
\includegraphics[width=0.24\textwidth]{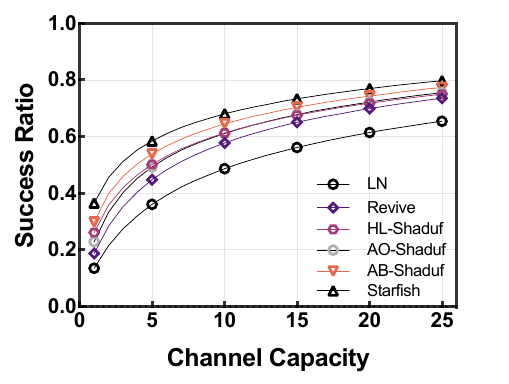}
\label{fig:Starfish Fixed Skew Small Payment}
}%
\subfigure[Uniform large payments]{
\includegraphics[width=0.24\textwidth]{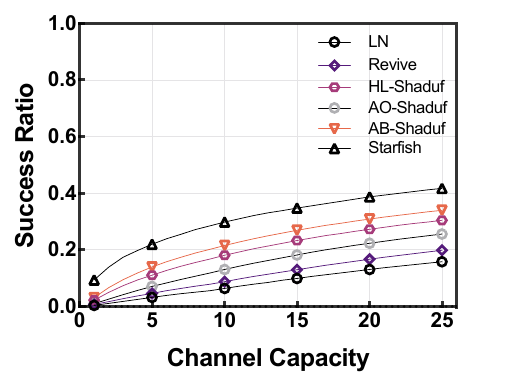}
\label{fig:Starfish Uniform large payment}
}%
\caption{Payment success ratios under different channel-capacity factors, payment skewness factors, and payment sizes. LN denotes the Lightning Network; HL, AO, and AB denote High-to-Low, All-to-One, and All-Bind, respectively.}
\label{fig:success ratio comparasion}
\end{figure*}

\textbf{The impact of channel capacity.}
From Figs.~\ref{fig:Starfish Uniform Small Payment}, \ref{fig:Starfish Uniform large payment}, and \ref{fig:Starfish Fixed Skew Small Payment}, it is apparent that as the channel capacity increases, the success rate of each protocol tends to rise. This is expected since a larger channel capacity enables the handling of a greater volume and size of transactions. However, notably, throughout the range of channel capacity from $1\times$ to $25\times$, Starfish consistently maintains a superior success ratio compared to other protocols.

\textbf{The impact of payment skewness.}
We explore the impact of payment skewness on the success ratio as shown in Fig.~\ref{fig:Starfish Skew Small Payment}. It can be observed that under different skewness levels, Starfish consistently exhibits the highest success ratio. Compared with LN's average success ratio, the average success ratios of Revive, HL-Shaduf, AO-Shaduf, AB-Shaduf, and Starfish are higher by around 9, 15, 14, 18, and 22 percentage points, respectively. We fix the payment skewness factor at 8 and evaluate the success ratio when channel capacities vary from $1\times$ to $25\times$. The results are presented in Fig.~\ref{fig:Starfish Fixed Skew Small Payment}. Compared with LN's average success ratio, the corresponding average success ratios are higher by around 8, 12, 12, 15, and 19 percentage points, respectively.

\textbf{The impact of payment size.}
We explore the impact of payment size on the success ratio. Figs.~\ref{fig:Starfish Uniform Small Payment} and \ref{fig:Starfish Uniform large payment} illustrate that Starfish consistently outperforms in both small and large payment scenarios. Under uniform small payments, the average success ratios of Revive, HL-Shaduf, AO-Shaduf, AB-Shaduf, and Starfish are higher than that of LN by around 8, 15, 14, 19, and 23 percentage points, respectively. Under uniform large payments, the corresponding differences are around 3, 11, 7, 15, and 23 percentage points, respectively.

% \subsection{Summary}
% \add{Shaduf's overall average success ratio surpasses LN and Revive, ranging from 4\% to 20\% and 2\% to 13\%, respectively. Starfish's overall average success ratio exceeds LN and Revive, ranging from 17\% to 26\% and 12\% to 19\%, respectively. AB-Shaduf's overall average success ratio surpasses HL-Shaduf and AO-Shaduf, ranging from 3\% to 5\% and 4\% to 6\%, but its gas cost significantly surpasses the latter two when involving a higher number of channels in rebalancing. In contrast, Starfish's overall average success ratio not only surpasses AB-Shaduf by 4\% to 7\% but also exhibits considerably lower gas costs than the latter when dealing with a larger number of channels in rebalancing, while also being lower than both HL-Shaduf and AO-Shaduf.}

\section{Conclusion}
\label{conslusion}
% In conclusion, our research introduces Starfish, a novel payment network addressing scalability challenges in blockchain Payment Channel Networks (PCNs). Starfish enhances rebalancing efficiency by allowing channels to borrow funds, demonstrating optimal performance in simulations. The Ethereum-based implementation proves practical feasibility. Compared to existing protocols, Starfish consistently outperforms in success ratios for off-chain payments, making it a promising solution for scalable and efficient blockchain transactions.

This paper presents Starfish, a rebalancing protocol designed for the locally star-like structures formed by routing hubs and their incident channels in real-world PCNs. By leveraging a hub-centric rebalancing mechanism, Starfish significantly reduces the on-chain overhead of rebalancing from quadratic to linear complexity, while maintaining high efficiency in off-chain payments. 
We formally establish the security of Starfish within the Universal Composability (UC) framework and validate its effectiveness through extensive simulations on realistic PCN topologies, as well as a proof-of-concept implementation on Ethereum. Experimental results show that Starfish consistently improves payment success ratios compared to existing rebalancing approaches, demonstrating its suitability for scalable and practical PCN deployments.

% Appendix A is included directly in this main source file.
% Full-width, non-breaking boxes used inside double-column floats.
\newtcolorbox{AppendixBox}[2][\linewidth]{%
    rounded corners, enhanced, colback=white,
    fontupper=\footnotesize, colframe=black, width=#1,
    toprule=1pt, rightrule=1pt, leftrule=1pt, bottomrule=1pt,
    left=1.5pt, right=1.5pt, bottom=1.5pt,
    fancytitle={#2}, pad at break=2mm}

\appendix

\section{Formal Security Model and Proof}\label{app:security-proof}
\subsection{Analysis of the Ideal Functionality}

\textbf{Analysis of the Ideal Functionality. }The complete specification of the ideal functionality $\mathcal{F}$ is presented in the full-width box associated with this appendix. In the following, we explain how $\mathcal{F}$ captures the desired security properties and summarize the corresponding round/time bounds.

% --- Begin: Ideal Functionality Box (without figure-style caption wrapper) ---
\begin{figure*}[!t]
\begin{AppendixBox}[\textwidth]{Ideal Functionality $\mathcal{F}$}
            
    % Assume the following messages concerning the channel $\beta$ and the merge $\varphi$, with their identifiers, $\beta.id$ and $\varphi.id$ denoted as $id$ and $\tilde{id}$, respectively. In the produces (B) and (D), we denote the requested payment and coin shift as $\theta$ and $\tilde{\theta}$, respectively. In the produce (E), we denote the coin shift between channels as $\hat{\theta}$. 
    
    \begin{center}
        (A) \textbf{Open Channel}
    \end{center}
    Upon receiving $(\mathtt{open}, \beta)$ from $A$ in $t$, let $B:=\beta.\mathsf{other\text{-}party}(A)$, proceed as follows: 
    \begin{enumerate}
        \item Within $t_1 \le  t + \Delta$, send $(\mathtt{remove},A,\beta.\mathsf{balanceC}(A))$ to ledger $\mathcal{L}$. 
        \item After completing step 1, send $(\mathtt{opening},\beta)$ to $B$ in $t_1$.
        \item Upon receiving $(\mathtt{open},\beta)$ from $B$ in $t_1$, send $(\mathtt{remove},B,\beta.\mathsf{balance}(B))$ to $\mathcal{L}$ within $t_2\le t_1+\Delta$, output $(\mathtt{opened})$ to $\beta.\mathsf{users}$ and simulator $\mathcal{S}$, then stop; otherwise, send $(\mathtt{add},A,\beta.\mathsf{balance}(A))$ to $\mathcal{L}$ after $t_2>t_1+\Delta$, output $(\mathtt{not\text{-}opened})$ to $A$, then stop. 
    \end{enumerate}
                            
    \begin{center}
        (B) \textbf{Update Channel}
    \end{center}
    Upon receiving $(\mathtt{updateC}, id, \theta)$ from $A$ in $t$, proceed as follows: 
    \begin{enumerate}
        \item Send $(\mathtt{updateC\text{-}req}, id, \theta)$ to $B$ in $t_1:=t+1$. To maintain synchronization with the rounds of our protocol, $\mathcal{F}$ waits for one round before sending the message. 
        \item Upon receiving $(\mathtt{updateC\text{-}ok})$ from $B$ in $t_1$, set $\beta.\mathsf{balance} := \beta.\mathsf{balance} + \theta$, output $(\mathtt{updated})$ to $A$ in $t_2:=t_1+1$, then stop; otherwise, output $(\mathtt{not\text{-}updated})$ to $A$, then stop. 
    \end{enumerate}

    \begin{center}
        (C) \textbf{Open Merge}
    \end{center}
    Upon receiving $(\mathtt{merge}, \varphi)$ from $\mathsf{hub}$ in $t$, proceed as follows:             
    \begin{enumerate}
        \item If $\mathsf{hub}$ is honest, send $(\mathtt{merge\text{-}req}, \varphi)$ to $P\in\varphi.\mathsf{users}$ in $t_1:=t+1$; otherwise, upon receiving $(\mathtt{send\text{-}req},P)$ from $\mathcal{S}$ in $t$, send $(\mathtt{merge\text{-}req}, \varphi)$ to $P$ in $t_1$. 
        \item Upon receiving $(\mathtt{merge},\varphi)$ from all users in $t_1$, send $(\mathtt{merge\text{-}confirm},\varphi)$ to $\mathsf{hub}$ in $t_2:=t_1+1$. 
        \item Upon receiving $(\mathtt{merge\text{-}confirmed},\varphi)$ from $\mathsf{hub}$ in $t_2$, let $\beta$ be the underlying channel for each edge $\epsilon\in\varphi.\mathsf{edges}$. Add $\varphi$ to $\beta.\mathsf{mergeSet}$. Set $\beta.\mathsf{balanceC}(\mathsf{hub}):=\beta.\mathsf{balanceC}(\mathsf{hub})-\varphi.\mathsf{capacity}(\epsilon)$. Output $(\mathtt{merged})$ to all users and $\mathcal{S}$ within $t_3\le t_2+\Delta$, then stop; otherwise, output $(\mathtt{not\text{-}merged})$, then stop.

    \end{enumerate}
    
    \begin{center}
        (D) \textbf{Update Edge}
    \end{center}
    
    Upon receiving $(\mathtt{updateE}, \tilde{id}, \tilde{\theta},\epsilon)$ from $\mathsf{hub}$ in $t$, proceed as follows: 
    \begin{enumerate}
        \item Send $(\mathtt{updateE\text{-}req}, \tilde{id}, \tilde{\theta})$ to $\mathsf{user}$ in $t_1:=t+1$.  
        \item Upon receiving $(\mathtt{updateE\text{-}ok})$ from $\mathsf{user}$ in $t_1$, output $(\mathtt{updatedE})$ to $\mathsf{hub}$ in $t_2:=t_1+1$, then stop; otherwise, output $(\mathtt{not\text{-}updatedE})$, then stop.
    \end{enumerate}

    \begin{center}
        (E) \textbf{Update Merge}
    \end{center}
    Let $\epsilon_1$ and $\epsilon_2$ be the edges involved, connecting to end-users $P$ and $Q$, respectively. Let $R\in\varphi.\mathsf{users}$. Upon receiving $(\mathtt{updateM}, \tilde{id}, \hat{\theta},\epsilon_1,\epsilon_2)$ from $\mathsf{hub}$ in $t$, proceed as follows: 
    \begin{enumerate}
        \item If $\mathsf{hub}$ is honest, send $(\mathtt{updateM\text{-}req}, \tilde{id}, \hat{\theta},\epsilon_1,\epsilon_2)$ to $P$ and $Q$ in $t_1:=t+1$; otherwise, upon receiving $(\mathsf{send\text{-}req},P)$ from $\mathcal{S}$ in $t$, send $(\mathtt{updateM\text{-}req}, \tilde{id}, \hat{\theta},\epsilon_1,\epsilon_2)$ to $P$ in $t_1$, and similarly for $Q$. 

        \item Upon receiving $(\mathtt{updateM\text{-}ok})$ from $P$ and $Q$ in $t_1$, send $(\mathtt{updateM\text{-}confirm}, \tilde{id}, \hat{\theta}, \epsilon_1,\epsilon_2)$ to $\mathsf{hub}$ in $t_2:=t_1+1$. 
        Upon receiving $(\mathtt{updateM\text{-}confirmed})$ from $\mathsf{hub}$ in $t_2$, send $(\mathtt{updateM\text{-}pending}, \tilde{id}, \hat{\theta}, \epsilon_1,\epsilon_2)$ to $\mathsf{users}$ in $t_3:=t_2+2$. 
        
        \item Upon receiving $(\mathtt{updateM\text{-}wrong})$ from $R$ and $R$ is honest in $t_3$, output $(\mathtt{not\text{-}updatedM})$ to $\varphi.\mathsf{users}$, and stop. Otherwise, update the $\mathsf{capacity}$ and $\mathsf{balanceE}$ of $\epsilon_1$ and $\epsilon_2$, respectively. Output $(\mathtt{updatedM})$ to $\varphi.\mathsf{users}$, and stop.
                        
    \end{enumerate}

    \begin{center}
        (F) \textbf{Close Merge}
    \end{center}

    Upon receiving $(\mathtt{closeM}, \tilde{id},\epsilon)$ from $\mathsf{hub}$ in $t$, proceed as follows: 
    \begin{enumerate}
        \item Assume $\mathsf{user}$ is the end party, remove $\mathsf{user}$ from $\varphi.\mathsf{users}$, remove $\epsilon$ from $\varphi.\mathsf{edges}$. 
        \item Let $\beta$ denote the underlying channel corresponding to $\epsilon$, remove $\varphi$ from $\beta.\mathsf{mergeSet}$, set $\beta.\mathsf{balanceC}:=\beta.\mathsf{balanceC}+\epsilon.\mathsf{balanceE}$. Output $(\mathtt{closedM}, \tilde{id},\epsilon)$ to $\varphi.\mathsf{users}$ and $\mathsf{user}$ within $t_1\le t+3\Delta$ and stop. 
    \end{enumerate}

    \begin{center}
        (G) \textbf{Close Channel}
    \end{center}

    Upon receiving $(\mathtt{closeC}, id)$ from $A$ in $t$, proceed as follows: 
    \begin{enumerate}
        \item If $\beta.\mathsf{mergeSet} = \emptyset$, within round $t_1 \le  t + 2\Delta$, send $(\mathtt{add},A,\beta.\mathsf{balanceC}(A))$ to ledger $\mathcal{L}$, and send ($\mathtt{add},B$, $\beta.\mathsf{balanceC}(B)$) to ledger $\mathcal{L}$. Output $(\mathtt{closedC})$ to $\beta.\mathsf{users}$, and stop.
        \item Otherwise, for each merge $\varphi$ in $\beta.\mathsf{mergeSet}$, execute procedure (F). Let $t_2$ be the current round. Within $t_3 \le  t_2 + 2\Delta$, send $(\mathtt{add},A,\beta.\mathsf{balanceC}(A))$ to ledger $\mathcal{L}$, and send $(\mathtt{add},B,\beta.\mathsf{balanceC}(B))$ to ledger $\mathcal{L}$. Output $(\mathtt{closedC})$ to $\beta.\mathsf{users}$, and stop.
    \end{enumerate}

\end{AppendixBox}
\end{figure*}
% --- End: Ideal Functionality Box ---

\balance
\noindent \textbf{Consensus on Open Channel and Update Channel. }A channel can only be opened when $A$ initiates the request and $B$ agrees, requiring a mutual consensus between the two parties. The time interval from the initiation of the open channel request to the actual opening is $2\Delta$. Similarly, an update of the channel also requires a consensus between both parties. Under the premise of both parties being honest, updating the channel takes 2 rounds.

\noindent \textbf{Consensus on Open Merge and Update Edge. }The procedure of open merge requires initiation by a hub and consent from all end users. The entire merging process is expected to take $\Delta+2$ rounds. The update edge process is similar to the update channel, necessitating consensus among the edge parties. If both parties remain honest, the operation is expected to take 2 rounds. 

\noindent \textbf{Consensus on Update Merge. }The update merge process mandates initiation by the hub, followed by confirmations from all the end users. In this procedure, the parties associated with the edges involved in the update merge are required to reach a consensus regarding the update merge request. Subsequently, the hub executes an atomic broadcast to synchronize the current update merge state. If all parties are honest, the procedure is expected to take 4 rounds. 

\noindent \textbf{Guaranteed Close Merge and Close Channel. }Within any merge edge, any party can issue a close merge request to remove its channel from the merge contract. The procedure is expected to be completed within $3\Delta$ time. Any party within a channel can issue a close channel request to close the channel. Should the channel not be integrated into a merge contract, it is expected to close within $2\Delta$ time; otherwise, it first requires $3\Delta$ time to close merge the channel before the closure can be executed. 

\noindent \textbf{Guaranteed balance payout for users. }Once a channel has been close merged, the current merge balance is transferred to the channel balance. When the channel is closed, the latest state of the channel balance is stored in the parties' account in the ledger $\mathcal{L}$. 

\FloatBarrier
\subsection{Simulation-Based Proof of Theorem~\ref{thm:uc-security}}

\begin{proof}
\renewcommand{\qedsymbol}{}
Simulator $\mathcal{S}$ facilitates interaction with the environment $\mathcal{E}$ and the ideal functionality $\mathcal{F}$. To achieve indistinguishability between the real and ideal worlds, $\mathcal{S}$ is required to simulate the behavior of the predefined adversary $\mathcal{A}$. This involves corrupting the same parties in the ideal world as adversary $\mathcal{A}$ does in the real world. $\mathcal{S}$ also needs to generate public-private key pairs for all parties, distributing the public keys to the corrupted parties. For each corrupted party, $\mathcal{S}$ sends the corresponding private key individually. $\mathcal{S}$ observes the actions of adversary $\mathcal{A}$ in the real world and the instructions given to the corrupted parties, then selects identical inputs to submit to the ideal functionality $\mathcal{F}$. Furthermore, $\mathcal{S}$ represents the contract function $\mathcal{C}$ and honest parties (i.e., dummy parties) to send messages to the corrupted parties. It is important to note that $\mathcal{S}$ does not simulate scenarios where all parties are either honest or corrupt. 
In cases involving two-party channels or edges, our simulation considers scenarios where either one party is honest or the other is corrupt. For multi-party interactions within the merge operations, we differentiate scenarios based on the hub being either honest or corrupt. Additionally, we simulate situations where the end users, according to this categorization, is portrayed as either honest or corrupt. 

\begin{figure*}[!t]
\begin{AppendixBox}[\textwidth]{Simulation}                       
            % Assume the following messages concerning the channel $\beta$ and the merge $\varphi$, with their identifiers, $\beta.\mathsf{id}$ and $\varphi.\mathsf{id}$ denoteld as $id$ and $\tilde{id}$, respectively. In the produces (B) and (D), we denote the requested payment and coin shift as $\theta$ and $\tilde{\theta}$, respectively. In the produce (E), we denote the coin shift between channels as $\hat{\theta}$, and we specify that the flow of merge balance is from channel $\beta$ to $\gamma$, $S$ is the end party of channel $\beta$, $R$ is the end party of channel $\gamma$, and $\mathsf{pSet}$ is the set of parties with priority greater than $R$ in $\varphi.\mathsf{priority}(\beta)$. We simplify $A=\beta.A$, $B=\beta.B$, $I=\varphi.\mathsf{hub}$, and denote the initiator of the channel update with $P$ and the responder with $Q$. Suppose the number of channels for merge is three.
            
            \begin{center}
                (A) \textbf{Open Channel} \\
                \textbf{Case: $A$ is honest and $B$ is corrupt}
            \end{center} 

            Upon $A$ sending $(\mathtt{open},\beta)$ to ideal functionality $\mathcal{F}$ in round $t$, send $(\mathtt{open},\beta)$ to contract instance $\mathcal{C}(\beta.id)$ on behalf of $A$ in the same round. Assume the message reaches the ledger $\mathcal{L}$ within $\tau\le t+\Delta$. If $B$ sends $(\mathtt{open},\beta)$ to $\mathcal{C}(\beta.id)$ in $\tau$, send $(\mathtt{open},\beta)$ to $\mathcal{F}$ on behalf of $B$ in the same round. 

            \begin{center}
                \textbf{Case: $A$ is corrupt and $B$ is honest}
            \end{center}

            Upon $A$ sending $(\mathtt{open},\beta)$ to contract instance $\mathcal{C}(\beta.id)$ in round $t$, send $(\mathtt{open},\beta)$ to ideal functionality $\mathcal{F}$ on behalf of $A$ in the same round. Assume the message reaches the ledger $\mathcal{L}$ within $\tau\le t+\Delta$. If $B$ sends $(\mathtt{open},\beta)$ to $\mathcal{F}$ in $\tau$, send $(\mathtt{open},\beta)$ to $\mathcal{C}(\beta.id)$ on behalf of $B$ in the same round.

            \begin{center}
                (B) \textbf{Update Channel} \\
                \textbf{Case: $A$ is honest and $B$ is corrupt}
            \end{center} 

            Upon $A$ sending $(\mathtt{updateC},id,\theta)$ to ideal functionality $\mathcal{F}$ in round $t$, sign $\sigma_A$ on $\mathsf{msgC}=(\beta^{(A)}.\mathsf{versionC}+1,\beta^{(A)}.\mathsf{balanceC}+\theta)$ and send $(\mathtt{updateC},id,\mathsf{msgC},\sigma_A)$ to $B$ on behalf of $A$ in the same round. If $B$ sends $(\mathtt{updateC},\sigma_B)$ to $A$ in $t_1:=t+1$, send $(\mathtt{updateC\text{-}ok})$ to $\mathcal{F}$ on behalf of $B$ in the same round. 

            \begin{center}
                \textbf{Case: $A$ is corrupt and $B$ is honest}
            \end{center}

            Let $\mathsf{msgC}=(v,b)$. Upon $A$ sending $(\mathtt{updateC},id,\mathsf{msgC},\sigma_A)$ to $B$ in round $t$, if $v=\beta^{(B)}.\mathsf{versionC}+1$, send $(\mathtt{updateC},id,b-\beta^{(B)}.\mathsf{balanceC})$ to $\mathcal{F}$ on behalf of $A$ in the same round; otherwise, ignore the message and stop. If $B$ sends $(\mathtt{updateC\text{-}ok})$ to $\mathcal{F}$ in $t_1:=t+1$, send $(\mathtt{updateC},\sigma_B)$ to $A$ on behalf of $B$ in the same round. 

            \begin{center}
                (C) \textbf{Open Merge} \\
                \textbf{Case: $\mathsf{hub}$ is honest}
            \end{center}

            Upon $\mathsf{hub}$ sending $(\mathtt{merge},\varphi)$ to ideal functionality $\mathcal{F}$ in round $t$, send $(\mathtt{merge},\varphi,\sigma_\mathsf{hub})$ to $P\in\varphi.\mathsf{users}$ on behalf of $\mathsf{hub}$ in the same round. Proceed as follows: 
            \begin{enumerate}
                \item If $P$ is corrupt and sends $(\mathtt{merge},\varphi,\sigma_P)$ to $\mathsf{hub}$ in $t_1:=t+1$, send $(\mathtt{merge},\varphi)$ to $\mathcal{F}$ on behalf of $P$ in the same round. If $P$ is honest and sends $(\mathtt{merge},\varphi)$ to $\mathcal{F}$ in $t_1$, send $(\mathtt{merge},\varphi,\sigma_P)$ to $\mathsf{hub}$ on behalf of $P$ in the same round.
                \item Upon $\mathsf{hub}$ sending $(\mathsf{merge\text{-}confirmed},\varphi)$ to $\mathcal{F}$ in $t_2:=t_1+1$, send $(\mathtt{merge},\varphi,\Sigma)$ to $\mathcal{C}(\varphi.\tilde{id})$ on behalf of $\mathsf{hub}$ in the same round, where $\Sigma$ represents the set of signatures from all users. 

            \end{enumerate}

            \begin{center}
                \textbf{Case: $\mathsf{hub}$ is corrupt}
            \end{center}

            Upon $\mathsf{hub}$ sending $(\mathtt{merge},\varphi,\sigma_\mathsf{hub})$ to $P\in\varphi.\mathsf{users}$ and $P$ is honest in round $t$, send $(\mathtt{merge},\varphi)$ to $\mathcal{F}$ on behalf of $\mathsf{hub}$ and send $(\mathsf{send\text{-}req},P)$ to $\mathcal{F}$ in the same round. Proceed as follows: 
            \begin{enumerate}
                \item If $P$ is corrupt, send $(\mathtt{merge},\varphi)$ to $\mathcal{F}$ on behalf of $P$ in $t_1:=t+1$. If $P$ is honest and sends $(\mathtt{merge},\varphi)$ to $\mathcal{F}$ in $t_1$, send $(\mathtt{merge},\varphi,\sigma_P)$ to $\mathsf{hub}$ on behalf of $P$ in the same round.
                \item Upon $\mathsf{hub}$ sending $(\mathtt{merge},\varphi,\Sigma)$ to $\mathcal{C}(\varphi.\tilde{id})$ in $t_2:=t_1+1$, where $\Sigma$ represents the set of signatures from all users, send $(\mathsf{merge\text{-}confirmed},\varphi)$ to $\mathcal{F}$ on behalf of $\mathsf{hub}$ in the same round. 
            \end{enumerate}

            \begin{center}
                (D) \textbf{Update Edge} \\
                \textbf{Case: $\mathsf{hub}$ is honest and $\mathsf{user}$ is corrupt}
            \end{center}

            Upon $\mathsf{hub}$ sending $(\mathtt{updateE},\tilde{id},\tilde{\theta},\epsilon)$ to ideal functionality $\mathcal{F}$ in round $t$, sign $\sigma_{\mathsf{hub}}$ on $\mathsf{msgE}=(\epsilon^{(\mathsf{hub})}.\mathsf{versionE}+1,\epsilon^{(\mathsf{hub})}.\mathsf{balanceE}+\tilde{\theta})$ and send ($\mathtt{updateE}$, $\tilde{id}$, $\mathsf{msgE}$, $\sigma_{\mathsf{hub}}$) to $\mathsf{user}$ on behalf of $\mathsf{hub}$ in the same round. If $\mathsf{user}$ sends $(\mathtt{updateE},\sigma_{\mathsf{user}})$ to $\mathsf{hub}$ in $t_1:=t+1$, send $(\mathtt{updateE\text{-}ok})$ to $\mathcal{F}$ on behalf of $\mathsf{user}$ in the same round. 

            \begin{center}
                \textbf{Case: $\mathsf{hub}$ is corrupt and $\mathsf{user}$ is honest}
            \end{center}
            
            Let $\mathsf{msgE}=(v,b)$, upon $\mathsf{hub}$ sending ($\mathtt{updateE}$, $\tilde{id}$, $\mathsf{msgE}$, $\sigma_{\mathsf{hub}}$) to $\mathsf{user}$ in round $t$, if $v=\epsilon^{(\mathsf{user})}.\mathsf{versionE}+1$, send $(\mathtt{updateE},\tilde{id},b-\epsilon^{(\mathsf{user})}.\mathsf{balanceE})$ to $\mathcal{F}$ on behalf of $\mathsf{hub}$ in the same round; otherwise, ignore the message and stop. If $\mathsf{user}$ sends $(\mathtt{updateE\text{-}ok})$ to $\mathcal{F}$ in $t_1:=t+1$, send $(\mathtt{updateE},\sigma_{\mathsf{user}})$ to $\mathsf{hub}$ on behalf of $\mathsf{user}$ in the same round.  

\end{AppendixBox}
\end{figure*}

\begin{figure*}[!t]
\begin{AppendixBox}[\textwidth]{Simulation (continued)}
            \begin{center}
                (E) \textbf{Update Merge} \\
                \textbf{Case: $\mathsf{hub}$ is honest}
            \end{center}

            Upon $\mathsf{hub}$ sending $(\mathtt{updateM}, \tilde{id},\hat{\theta},\epsilon_1,\epsilon_2)$ to ideal functionality $\mathcal{F}$ in round $t$, sign $\sigma_{\mathsf{hub}}$ on $\mathsf{msgM}=$ ($\varphi^{(\mathsf{hub})}.\mathsf{versionM}+1$, $\mathsf{cap}^{(\mathsf{hub})}$) and $\mathsf{msgE}=$ ($\epsilon_1^{(\mathsf{hub})}.\mathsf{versionE}+1$, $\epsilon_2^{(\mathsf{hub})}.\mathsf{versionE}+1$, $\epsilon^{(\mathsf{hub})}.\mathsf{capacity}+\hat{\theta}$, $\epsilon^{(\mathsf{hub})}.\mathsf{balanceE}(\mathsf{hub})+\hat{\theta}$), send $(\mathtt{updateM},\tilde{id},\mathsf{msgM},\mathsf{msgE},\sigma_{\mathsf{hub}})$ to $P$ and $Q$ (the end-users of $\epsilon_1$ and $\epsilon_2$, respectively) on behalf of $\mathsf{hub}$ in the same round. Proceed as follows: 

            \begin{enumerate}
                \item If $P$ (or $Q$) is corrupt and sends $(\mathtt{updateM},\sigma_P)$ to $\mathsf{hub}$ in $t_1:=t+1$, send $(\mathtt{updateM\text{-}ok})$ to $\mathcal{F}$ on behalf of $P$ (or $Q$) in the same round. 
                If $P$ (or $Q$) is honest and sends $(\mathtt{updateM}\text{-}\mathtt{ok})$ to $\mathcal{F}$ in $t_1$, send $(\mathtt{updateM},\sigma_P)$ to $\mathsf{hub}$ on behalf of $P$ (or $Q$) in the same round. 

                \item If $\mathsf{hub}$ sends $(\mathtt{updateM\text{-}confirmed})$ to $\mathcal{F}$ in $t_2:=t_1+1$, run $\mathsf{AtomicBroadcast}(\mathtt{updateM},\tilde{id},\mathsf{msgM},\mathsf{msgE},\Sigma)$ to $\varphi.\mathsf{users}$ on behalf of $\mathsf{hub}$ in the same round, where $\Sigma$ collects the above signatures. 

            \end{enumerate}

            \begin{center}
                \textbf{Case: $\mathsf{hub}$ is corrupt}
            \end{center}

            Let $\mathsf{msgM}=(vM,c)$ and $\mathsf{msgE}=(vE_1,vE_2,\hat{c},\hat{b})$, upon $\mathsf{hub}$ sending $(\mathtt{updateM},\tilde{id},\mathsf{msgM},\mathsf{msgE},\sigma_{\mathsf{hub}})$ to $P$ (or $Q$) and $P$ is honest in round $t$, send $(\mathtt{updateM}, \tilde{id},\hat{\theta},\epsilon_1,\epsilon_2)$ to $\mathcal{F}$ on behalf of $\mathsf{hub}$ and send $(\mathtt{send\text{-}req},P)$ to $\mathcal{F}$ in the same round. Proceed as follows: 

            \begin{enumerate}
                \item If $P$ (or $Q$) is corrupt, send $(\mathtt{updateM}\text{-}\mathsf{ok})$ to $\mathcal{F}$ on behalf of $P$ in $t_1:=t+1$. 
                If $P$ is honest and sends $(\mathtt{updateM}\text{-}\mathsf{ok})$ to $\mathcal{F}$ in $t_1$, send $(\mathtt{updateM},\sigma_P)$ to $\mathsf{hub}$ on behalf of $P$ in the same round. 

                \item If $\mathsf{hub}$ runs $\mathsf{AtomicBroadcast}(\mathtt{updateM},\tilde{id},\mathsf{msgM},\mathsf{msgE},\Sigma)$ to $\varphi.\mathsf{users}$ in $t_2:=t_1+1$, where $\Sigma$ collects the above signatures. Send $(\mathtt{updateM\text{-}confirmed})$ to $\mathcal{F}$ on behalf of $\mathsf{hub}$ in the same round. 

                \item If $R\in\varphi.\mathsf{users}$ is honest, and sends $(\mathtt{updateM\text{-}wrong})$ to $\mathcal{F}$ in $t_3:=t_2+2$, ignore the $\mathsf{AtomicBrodcast}$ messages on behalf of $R$. 
                
            \end{enumerate}

            \begin{center}
                (F) \textbf{Close Merge} \\
                \textbf{Case: $\mathsf{hub}$ is honest and $\mathsf{user}$ is corrupt}
            \end{center}

            Upon $\mathsf{hub}$ sending $(\mathtt{closeM},\tilde{id},\epsilon)$ to ideal functionality $\mathcal{F}$ in round $t$, set $\mathsf{msgM}=$ ($\varphi^{(\mathsf{hub})}.\mathsf{versionM}$, $\mathsf{cap}^{(\mathsf{hub})}$, $\Sigma_R$), where $\Sigma_R$ is the signature set of the current merge state. Set $\mathsf{msgE}=$ ($\epsilon^{(\mathsf{hub})}.\mathsf{versionE}$, $\epsilon^{(\mathsf{hub})}.\mathsf{balanceE}$, $\Sigma_E$), where $\Sigma_E$ is the signature set of the current edge state. Send ($\mathtt{closeM}$, $\tilde{id}$, $\epsilon$, $\mathsf{msgM}$, $\mathsf{msgE}$) to $\mathcal{C}(\varphi.\tilde{id})$ on behalf of $\mathsf{hub}$ in the same round. Proceed as follows: 
            \begin{enumerate}
                \item Assume the message reaches ledger $\mathcal{L}$ within round $\tau\le t + \Delta$. If $\mathsf{user}$ does not send any message, send $(\mathtt{timeout},\tilde{id})$ to $\mathcal{C}(\varphi.\tilde{id})$ on behalf of $\mathsf{hub}$ in round $\tau_1>\tau+\Delta$.
                \item Let $R\in \varphi.\mathsf{users}\backslash\{\mathsf{user}\}$. If $R$ is corrupt and does not send message to $\mathcal{C}(\varphi.\tilde{id})$ within round $\tau_2\le \tau+2\Delta$. Send confirmation to $\mathcal{F}$ on behalf of $R$ in the same round. 

            \end{enumerate}

            \begin{center}
                \textbf{Case: $\mathsf{hub}$ is corrupt and $\mathsf{user}$ is honest}
            \end{center}

            Upon $\mathsf{hub}$ sending ($\mathtt{closeM}$, $\tilde{id}$, $\epsilon$, $\mathsf{msgM}$, $\mathsf{msgE}$) to $\mathcal{C}(\varphi.\tilde{id})$ in round $t$, send $(\mathtt{closeM},\tilde{id},\epsilon)$ to ideal functionality $\mathcal{F}$ on behalf of $\mathsf{hub}$ in the same round. Proceed as follows: 
            \begin{enumerate}
                \item Assume the message reaches ledger $\mathcal{L}$ within round $\tau\le t + \Delta$, send ($\mathtt{closeM}$, $\tilde{id}$, $\epsilon$, $\mathsf{msgM}$, $\mathsf{msgE}$) to $\mathcal{C}(\varphi.\tilde{id})$ on behalf of $\mathsf{user}$ in $\tau$. 
                \item Let $R\in \varphi.\mathsf{users}\backslash\{\mathsf{user}\}$. If $R$ is honest and sends confirmation to $\mathcal{F}$ within round $\tau_2\le \tau+2\Delta$, send ($\mathtt{closeM\text{-}challenge}$, $\tilde{id}$, $\mathsf{msgM}$) to $\mathcal{C}(\varphi.\tilde{id})$ on behalf of $R$ in the same round.

            \end{enumerate}

            \begin{center}
                (G) \textbf{Close Channel} \\
                \textbf{Case: $A$ is honest and $B$ is corrupt}
            \end{center}

            Upon $A$ sending $(\mathtt{closeC},id)$ to ideal functionality $\mathcal{F}$ in round $t$, generates $\mathsf{msgC}= $ ($\beta^{(A)}.\mathsf{versionC}$, $\beta^{(A)}.\mathsf{balanceC}$, $\Sigma_C$), where $\Sigma_C$ denotes the signatures of $A$ and $B$ for the current channel state. Sends ($\mathtt{closeC}$, $id$, $\mathsf{msgC}$) to $\mathcal{C}(\beta.id)$ on behalf of $A$ in the same round. Assume the message reaches ledger $\mathcal{L}$ within round $\tau\le t + \Delta$. If $\beta.\mathsf{mergeSet}=\emptyset$ and $B$ sends ($\mathtt{closeC}$, $id$, $\mathsf{msgC}$) to $\mathcal{C}(\beta.id)$ in round $\tau$, where $\mathsf{msgC}= $ ($\beta^{(B)}.\mathsf{versionC}$, $\beta^{(B)}.\mathsf{balanceC}$, $\Sigma_C$), sends confirmation to $\mathcal{F}$ on behalf of $B$ in the same round. Otherwise, for each merge in $\beta.\mathsf{mergeSet}$, $A$ executes Procedure (F) in round $\tau$.

            \begin{center}
                \textbf{Case: $A$ is corrupt and $B$ is honest}
            \end{center}

            Upon $A$ sending ($\mathtt{closeC}$, $id$, $\mathsf{msgC}$) to $\mathcal{C}(\beta.id)$ in round $t$, where $\mathsf{msgC}= $ ($v$, $b$, $\Sigma_C$), sends $(\mathtt{closeC},id)$ to ideal functionality $\mathcal{F}$ on behalf of $A$ in the same round. Assume the message reaches ledger $\mathcal{L}$ within round $\tau\le t + \Delta$. If $\beta.\mathsf{mergeSet}=\emptyset$, sends ($\mathtt{closeC}$, $id$, $\mathsf{msgC}$) to $\mathcal{C}(\beta.id)$ on behalf of $B$ in round $\tau$, where $\mathsf{msgC}= $ ($\beta^{(B)}.\mathsf{versionC}$, $\beta^{(B)}.\mathsf{balanceC}$, $\Sigma_C$). Otherwise, for each merge in $\beta.\mathsf{mergeSet}$, $B$ executes Procedure (F) in round $\tau$.  
        
\end{AppendixBox}
\end{figure*}
\end{proof}
\FloatBarrier

% Flush appendix floats before the two-column back matter.
\clearpage

\printcredits

\section*{Declaration of Competing Interest}
The authors declare that they have no known competing financial interests or personal relationships that could have appeared to influence the work reported in this paper.

\section*{Declaration of generative AI and AI-assisted technologies in the manuscript preparation process}
The authors used OpenAI Codex to assist with converting the manuscript to the LaTeX format and refining its presentation. All protocol designs, algorithms, security arguments, implementation logic, and experimental results were developed by the authors. All AI-assisted formatting changes were manually reviewed and edited by the authors, who take full responsibility for the content of the publication.

\section*{Acknowledgment}
\begin{sloppypar}
This study was supported by the National Natural Science Foundation of China (No. U23A20302, U25A20425, 62302266, 62232010, U24A20244), project ZR2026QA06 supported by Shandong Provincial Natural Science Foundation, Taishan Scholars Program under Grant tsqn202507069, Key technology research and industrialization demonstration projects in Qingdao 26-1-1-gjgg-30-gx and the Research Project of Quancheng Laboratory, China under Grant No. QCL20250106.
\end{sloppypar}

\bibliographystyle{cas-model2-names}
\balance
\bibliography{references}
\end{document}